\documentclass[aps,prb,twocolumn,amsfonts,amsmath,amssymb,superscriptaddress,showpacs]{revtex4-1}

\usepackage{graphicx}
\usepackage[utf8]{inputenc}
\usepackage{subfigure}
\usepackage{listings}
\usepackage{bm}
\usepackage{setspace}
\usepackage{hyperref}
\hypersetup{colorlinks=true,citecolor=blue,urlcolor=blue,linkcolor=blue,hyperfigures=true}
\usepackage{natbib}
\usepackage[normalem]{ulem}
\usepackage{epstopdf}
\usepackage{subfigure}
\usepackage{color}
\usepackage{dsfont}
\usepackage[bbgreekl]{mathbbol}
\usepackage{amsfonts}
\usepackage{amsmath, amssymb}
\usepackage{amsthm}
\usepackage{soul}

\newcommand{\dn}{\downarrow}
\newcommand{\up}{\uparrow}

\makeatother
\begin{document}

\title{Interplay between magnetic and superconducting fluctuations in the doped 2D Hubbard model:  {a parquet equations study}}

\author{Anna Kauch}
\affiliation{Institute for Solid State Physics, Vienna University of Technology, A-1040 Vienna, Austria}

\author{Felix H\"orbinger}
\affiliation{Institute for Solid State Physics, Vienna University of Technology, A-1040 Vienna, Austria}
\author{Motoharu Kitatani}
\affiliation{Institute for Solid State Physics, Vienna University of Technology, A-1040 Vienna, Austria}
\affiliation{RIKEN Center for Emergent Matter Sciences, Wako, Saitama, 351-0198, Japan}
\author{Gang Li}
\affiliation{School of Physical Science and Technology, ShanghaiTech University, Shanghai 201210, China}
\author{Karsten Held}
\affiliation{Institute for Solid State Physics, Vienna University of Technology, A-1040 Vienna, Austria}

\date{\today}

\begin{abstract}
We study the Hubbard model on a square lattice, using the  {parquet} dynamical vertex approximation and the parquet approximation. These methods allow us to describe the mutual interference  of spin-fluctuations in the particle-hole channel and superconducting fluctuations in the cooperon channel in an unbiased way. For small dopings we find predominant commensurable antiferromagnetic spin- and $d$-wave superconducting fluctuations; for larger doping incommensurate  antiferromagnetic spin fluctuations are concomitant to  spin-triplet odd-frequency $s$-wave superconducting fluctuations which are also supported by the dynamical structure of the pairing vertex.

\end{abstract}


\maketitle


\section{Introduction}

Unconventional superconductivity is not conventional in many respects.  First of all, the critical temperatures are often high, which means several ten up to one hundred Kelvin. Second, the materials such as cuprates, \cite{Bednorz1986,Dagotto1994} ruthenates,\cite{Maeno1994,Mackenzie2003} and iron pnictides \cite{Kamihara2006,Wang2008} are special: they belong to the class of strongly correlated electron systems where a strong Coulomb interaction prevents a simple one-particle description. Correlations between the electrons need to be considered. Third, the symmetry of the order parameter is rather unusual. Instead of a simple $s$-wave symmetry as in conventional, phonon-based superconductors, $d$-\cite{Wollman1993,Shen1993,Tsuei1994,Maki1998},  $p$-\cite{Maeno1994,Sigrist1999,Mackenzie2003} or $s_{\pm}$-wave symmetries\cite{Mazin2008b,Kuroki2008,Ikeda2010,Hanaguri2010,Hirschfeld2011,Kalenyuk2018} have been suggested and observed. 

The latter two facts are strongly related to each other. In contrast to the  phonon-mediated conventional superconductivity, unconventional superconductors are considered to occur due to the repulsive Coulomb interaction. Therefore even if spin fluctuations~\cite{Scalapino_RMP} or other microscopic mechanisms generate some attraction between the electrons, it cannot be local and instantaneous. The consequence is unconventional symmetry of the order parameter instead of a plain vanilla (even-frequency) $s$-wave.

The theoretical description of  high-temperature superconductivity  has been---and still is---a challenge. It is pretty much established by now that one of the fundamental models for  high-temperature superconductivity, the Hubbard model, indeed shows antiferromagnetism, pseudogaps, and $d$-wave superconductivity.
This has been demonstrated by a number of complementary methods such as quantum Monte-Carlo simulations,\cite{White1989,White1989b,Bulut1993,Scalapino1999,bulut} the functional-renormalization group (fRG), \cite{Zanchi1998,Halboth2000,Halboth2000b,Honerkamp2001c,Honerkamp2001b,Metzner2012}, the fluctuation-exchange approximation (FLEX),~\cite{Bickers1989,Moriya1990,Manske2004}, the variational cluster approximation \cite{Hanke2006} the two-particle self-consistent  theory, \cite{Kyung2003} cluster \cite{Lichtenstein2000,Maier2005,Maier2005a,Capone2006,Haule2007b,Sordi2012,Gull2013} and diagrammatic extensions \cite{Toschi2007,Rubtsov2009,Katanin2009,Hafermann2009a,Otsuki2014,Taranto2014,Kitatani2015,Kitatani2017,Ayral2015,Gukelberger2016,Vucicevic2017,Kitatani2018,RMPVertex} of dynamical mean field theory (DMFT). In the far overdoped region,  $d$-wave superconductivity is not stable any longer.  Since an interacting fermion system is however prone to all kinds of instabilities at low temperatures, one expects another kind of magnetic or superconducting instability at larger doping. One out of several possible candidates is $s$-wave superconductivity, which has been studied to a limited extent in Refs. \onlinecite{White1989b,Aligia1998,Maier_RPA,Kitatani2017}.

In the present paper, we study the two-dimensional Hubbard model on a square lattice, employing  the parquet approximation (PA) \cite{Diatlov1957,DeDominicis1962,DeDominicis1964} and the dynamical vertex approximation (D$\Gamma$A)\cite{Toschi2007,RMPVertex} in its parquet variant.\cite{Valli2015,Li2016,Pudleiner2019,Pudleiner2019a,Kauch2019}  {This is the first study of a lattice model out of half filling with the parquet variant of D$\Gamma$A}.  In contrast to previous diagrammatic extensions of DMFT, \cite{Rubtsov2009,Katanin2009,Hafermann2009a,Otsuki2014,Taranto2014,Kitatani2015,Ayral2015,Gukelberger2016,Vucicevic2017,Kitatani2018,Sayyad2019} this allows us to include the mutual feedback of the particle-hole and particle-particle channel, and hence represents a more  rigorous treatment
of the interplay between antiferromagnetic and superconducting fluctuations. All instabilities (magnetic, superconducting, charge density wave etc.) are treated on an equal footing. In the presence of several competing instabilities and their mutual feedback, a method that is not biased in favor (or against) a certain channel  such as the PA and parquet D$\Gamma$A is required. We are not only unbiased with respect to the channels but also do not restrict the available symmetries by introducing formfactors or neglecting the frequency dependence, as done in many fRG calculations. 
The downside is a huge numerical effort that restricts available temperatures $T$ and interaction strengths $U$. In the accessible $T$-range, we are able to identify the leading magnetic and superconducting fluctuations which  {could} lead to a corresponding order at  lower $T$'s.  {Although we cannot conclude which instability will dominate at low temperatures, we see a suppression of $d$-wave pairing fluctuations with doping and occurence of odd-frequency $s$-wave fluctuations in the triplet pairing channel. }   

For small doping levels (filling around $n=0.85$), we find commensurate antiferromagnetism with wavevector  ${\mathbf q}= (\pi,\pi)$ and $d$-wave superconducting fluctuations.  We confirm that the antiferromagnetic fluctuations and through them the superconducting fluctuations are strongly suppressed if one takes the full dynamics of the vertex into account. In the parquet approaches (both D$\Gamma$A and PA) this suppression is even stronger than in ladder  D$\Gamma$A.\cite{Kitatani2018}
For larger doping (filling around $n=0.75$),  the dominant magnetic wave vector becomes incommensurate with ${\mathbf q}_{1/2}= (\pi\pm\delta,\pi)$, $\delta\neq 0$, cf.~Refs.~\onlinecite{Schulz1990,Freericks1995,Jarrell1997,Yamase2016,Schaefer2016,Kozik_PRB_2017}, accompanied by triplet $s$-wave superconducting fluctuations. 
 This triplet $s$-wave symmetry implies an order parameter that is odd in frequency~\cite{Balatsky_review} and, as we will show, it originates from the frequency structure of the vertex.  Hitherto,  such a pairing  was found for a frustrated 2D~\cite{Vojta1999,Tanaka2008} and quasi-1D lattice~\cite{Tanaka2009}, as well as in staggered field~\cite{Tanaka2012}, in Josephson junctions~\cite{Balatsky_review} and Kondo lattices~\cite{Otsuki2015}. But to the best of our knowledge it has not been discussed before for the unfrustrated 2D Hubbard model.

The paper is organized as follows: We start with a brief introduction of the model in Section \ref{Sec:Model} and of the method  in Section \ref{Sec:Method}. The latter includes a recapitulation of the parquet equations (Section \ref{Sec:parquet})
taking the bare interaction (Section \ref{Sec:ParquetApprox}; PA) or the local fully irreducible vertex as input (Section \ref{Sec:ParquetDGA}, parquet D$\Gamma$A). Section \ref{Sec_eigenvalues}  further outlines how to use the eigenvalues of the Bethe-Salpeter equation as indicators for symmetry breaking;  and Section \ref{Sec_numerical} introduces improvements of the {\em victory} code
 that utilize the point group symmetry for reducing the number of $\mathbf k$-points to be considered and a coarse graining 
that employs a finer $\mathbf k$-grid for the Green's function bubbles.

Section \ref{Sec:Results} presents the results obtained, starting with
the analysis  of the  eigenvalues of the Bethe-Salpeter equation
in Section \ref{Sec:DominantEW}. Commensurate antiferromagnetic and $d$-wave superconducting fluctuations dominate for smaller doping  whereas incommensurate antiferromagnetism and  triplet $s$-wave  {pairing fluctuations} have the largest eigenvalues for large dopings. The temperature dependence of the eigenvalues in Section \ref{Sec:Tdep} shows that  $d$-wave superconductivity is likely for lower temperatures, even though a definite conclusion is hampered by the limited temperature interval available; for the  triplet $s$-wave superconducting a reliable prediction is not possible given the  temperature range accessible. In Section \ref{Sec:DynVertex} we present the dynamics of the vertex for the two different dopings. The symmetries  of the  $d$-wave   (Section \ref{dwave}) and triplet $s$-wave eigenvectors (Section \ref{Sec:swave}) are analyzed in $\mathbf k$ and real space. In Section \ref{Sec:swave}, we also trace back the origin of the  $s$-wave superconducting fluctuations;  and in  Section \ref{sec:ladder} we compare with ladder D$\Gamma$A results.
We sum up our results  in Section \ref{Sec:Summary} and provide further information on the convergence with respect to the number of frequencies in Appendix \ref{Apppendix:convergence}.

\section{Model}
\label{Sec:Model}

As a simple, minimal model for the physics of cuprates we consider the Hubbard model on a square lattice, mimicking the $d_{x^2-y^2}$ Cu orbitals of the CuO$_2$ layers but neglecting an explicit treatment of the oxygen $p$ orbitals:
\begin{equation}
\mathcal{H} = -t\sum_{\langle ij \rangle,\sigma} c_{i\sigma}^{\dagger}c_{j\sigma}^{\phantom{\dagger}} + U\sum_{i}n_{i\uparrow}n_{i\downarrow}.
\label{eq:Hubbard}
\end{equation}
Here, $c_{i\sigma}^{\dagger}$ ($c_{i\sigma}$) creates (annihilates) an electron on site $i$ with spin $\sigma$; $\langle ij \rangle$ denotes the summation over nearest neighbors only, $U$ is the  local Coulomb repulsion and $t$ the hopping amplitude. In the following, all energies are measured in units of  $t\equiv 1$.

The Hubbard model has been studied intensively in the context of high-temperature superconductivity. Initially  the infamous sign problem prevented a clear numerical answer from quantum Monte-Carlo simulations on whether this model describes $d$-wave superconductivity or not. But by now there is compelling evidence that the Hubbard model indeed describes  $d$-wave superconductivity akin to the cuprates. This understanding is supported by different many-body calculations, such as the functional renormalization group approach \cite{Metzner2012}, the fluctuation exchange approximation \cite{Bickers1989} as well as the application of quantum Monte-Carlo simulations \cite{Scalapino2007}, cluster extensions of DMFT  \cite{Lichtenstein2000,Maier2005a,Capone2006,Gull2013} and diagrammatic extensions of DMFT \cite{Otsuki2014,Kitatani2018}. Note that the latter adopted a simplified ladder approach instead of the more complete parquet equations employed here, which is crucial for faithfully treating the competition between different fluctuations. 


\section{Method}
\label{Sec:Method}

\subsection{Parquet equations}

Let us start by recapitulating the so-called parquet equations, following Ref.~\onlinecite{victory}

\label{Sec:parquet}
The parquet formalism \cite{Diatlov1957,DeDominicis1962,DeDominicis1964,Janis2001,Janis2017} consists of a set of exact equations
that connect the  single-particle Green's function and self-energy with two-particle vertex functions:
{i.e.}\ the Dyson equation, the Schwinger-Dyson equation, the Bethe-Salpeter equations in all channels, and the actual parquet equation.
The essential concept is to classify Feynman diagrams in terms of their reducibility. A diagram is called one-(two-)particle reducible if it falls into two pieces when cutting one(two) Green's function lines.

On the one-particle level the Green's function $G$ contains all diagrams (reducible or not), whereas the self-energy $\Sigma$ contains all one-particle irreducible~\cite{footnote1} diagrams with one incoming and one outgoing particle (leg)
in terms of the bare, non-interacting Green's function $G_0$.

 To be self-contained and at the same time to keep the discussion simple,  let us first introduce the necessary functions which contain all topologically invariant one- and two-particle Feynman diagrams.
We start with the one-particle functions: 
 From  $\Sigma$ all (reducible and irreducible) Green's function $G$ diagrams are obtained through 
 the Dyson equation 
\begin{equation}\label{GF}
G_k = [G^{-1}_{0,k} - \Sigma_k]^{-1} = \left[ i\nu + \mu - \epsilon_{\mathbf{k}} - \Sigma_k\right]^{-1}\;,
\end{equation}
with the chemical potential $\mu$, the fermionic Matsubara frequency $\nu$, and a  combined notation of momentum $\mathbf{k}$ and frequency $\nu$ in form of a four-vector, {i.e.} $k=(\mathbf{k}, \nu)$. When the self-energy is zero, one obtains the non-interacting Green's function $G_{0,k}=( i\nu + \mu - \epsilon_{\mathbf{k}} )^{-1}$ determined entirely by the dispersion relation $\epsilon_{\mathbf{k}}$. Please note, that if we express the self-energy in terms of Feynman diagrams with the interacting Green's function $G$ instead of $G_0$  only two-particle irreducible skeleton diagrams must be considered (all the other diagrams are generated through self-energy inclusions, i.e., through Eq.~\eqref{GF}). 

\begin{figure}[tb]
\centering
\includegraphics[width=\linewidth]{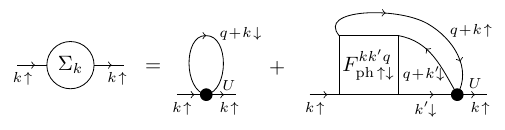}
\caption{Feynman diagrammatic representation of the Schwinger-Dyson  Eq.~(\ref{DysSchw}), where the identity $F_{\!ph\up\!\dn}^{k,k^{\prime},q}=\frac{1}{2}[F^{k,k^{\prime},q}_d-F^{k,k^{\prime},q}_m]$ is used.}
\label{fig:DysSchw}
\end{figure}

But we do not need to explicitly consider all of these diagrams. In the parquet formalism, the self energy is obtained from the full two-particle vertex $F$ through the Schwinger-Dyson equation also known as (Heisenberg) equation of motion
(cf.~Fig.~\ref{fig:DysSchw})
\begin{align}\label{DysSchw}
\Sigma_k = \frac{Un}{2}-\frac{U}{2}\sum_{k^{\prime},q}[F_{d}^{k,k^{\prime},q}-F_{m}^{k,k^{\prime},q}]G_{k+q}G_{k^{\prime}}G_{k^{\prime}+q}\;,
\end{align}
where $n$ is the number of electrons per site, and we have implicitly assumed a proper normalization of the momentum and frequency sums $\sum_{k}=1$, i.e., a prefactor $T$/(number of ${\mathbf k}$ points) is included in the definition of the sum.

As for the two-particle quantities the parquet equation classifies 
the full vertex functions $F$  further: namely into the fully two-particle irreducible class of diagrams $\Lambda$ and classes of diagrams $\Phi_r$ which are reducible regarding a specific channel $r\in\{ph,\overline{ph},pp\}$, i.e., particle-hole, transversal particle-hole and particle-particle channel.
For  $\mbox{SU}(2)$-symmetry it is convenient to further use some combinations of the  spin indices known as the density ($d$), magnetic ($m$), singlet ($s$) and triplet ($t$) channel (in addition to the   $r$ channels regarding the irreducibility).
This leads to the following parquet equations in four independent channels, see~\cite{victory,Li2016,Bickers-Review,RMPVertex}:
\begin{subequations}\label{PA_F}
\begin{align}
F_{d}^{k,k^{\prime},q}& =\Lambda_{d}^{k,k^{\prime},q} + \Phi^{k,k^{\prime},q}_{d}- \frac{1}{2}\Phi_{d}^{k,k+q,k^{\prime}-k} - \frac{3}{2}\Phi_{m}^{k,k+q,k^{\prime}-k} \nonumber \\
&+\frac{1}{2}\Phi_{s}^{k,k^{\prime},k+k^{\prime}+q} +\frac{3}{2}\Phi_{t}^{k,k^{\prime},k+k^{\prime}+q} \; ,\\
F_{m}^{k,k^{\prime},q}& =\Lambda_{m}^{k,k^{\prime},q} + \Phi^{k,k^{\prime},q}_{m} - \frac{1}{2}\Phi_{d}^{k,k+q,k^{\prime}-k}
+ \frac{1}{2}\Phi_{m}^{k,k+q,k^{\prime}-k} \nonumber \\
& -\frac{1}{2}\Phi_{s}^{k,k^{\prime},k+k^{\prime}+q}
+\frac{1}{2}\Phi_{t}^{k,k^{\prime},k+k^{\prime}+q}\; ,\\
F_{s}^{k,k^{\prime},q}& =\Lambda_{s}^{k,k^{\prime},q} + \Phi^{k,k^{\prime},q}_{s}
 +\frac{1}{2}\Phi_{d}^{k,q-k^{\prime},k^{\prime}-k}
 -\frac{3}{2}\Phi_{m}^{k,q-k^{\prime},k^{\prime}-k}\nonumber \\
& +\frac{1}{2}\Phi_{d}^{k,k^{\prime},q-k-k^{\prime}}
 - \frac{3}{2}\Phi_{m}^{k,k^{\prime},q-k-k^{\prime}}\; ,\\
F_{t}^{k,k^{\prime},q} & = \Lambda_{t}^{k,k^{\prime},q}  + \Phi^{k,k^{\prime},q}_{t}
 -\frac{1}{2}\Phi_{d}^{k,q-k^{\prime},k^{\prime}-k}
 -\frac{1}{2}\Phi_{m}^{k,q-k^{\prime},k^{\prime}-k} \nonumber \\
&+\frac{1}{2}\Phi_{d}^{k,k^{\prime},q-k-k^{\prime}}
+ \frac{1}{2}\Phi_{m}^{k,k^{\prime},q-k-k^{\prime}}\:. 
\end{align}
\end{subequations} 
Here, $F_{d/m/s/t}^{k,k^{\prime},q}$ are the complete vertices in the corresponding channels (combining spin indices and $r$ channels). Note that we use $d/m$ (density/magnetic) subscripts in place of $c/s$ (charge/spin) employed in Ref.~\onlinecite{RMPVertex} to avoid confusion with the  particle-particle singlet channel.

\begin{figure}[tb]
\centering
\includegraphics[width=\linewidth]{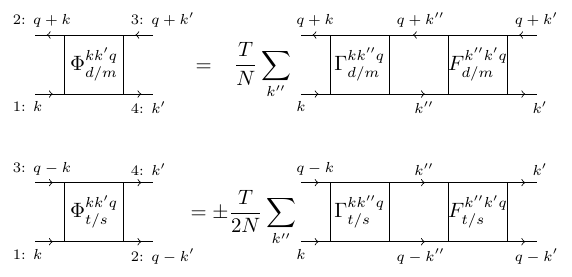}
\caption{ Feynman diagrammatic representation of the Bethe-Salpeter equations (\ref{PA_F_Phi}). }
\label{fig:PA_F_Phi}
\end{figure}

The vertex functions $\Phi_{d/m}^{k,k^{\prime},q}$ and $\Phi_{s/t}^{k,k^{\prime},q}$ in Eqs.~\eqref{PA_F}  denote the reducible vertex in the particle-hole ($d/m$) and particle-particle ($s/t$) channel, respectively.
They can be calculated via the Bethe-Salpeter equation in the respective channel, i.e., a ladder with the  irreducible vertex function $\Gamma^{k,k^{\prime},q}_{d/m/s/t} \equiv F_{d/m/s/t}^{k,k^{\prime},q} - \Phi_{d/m/s/t}^{k,k^{\prime},q}$ in the respective channel and  two connecting Green's functions
as building blocks. The  Bethe-Salpeter equations can be cast into the following form, cf.\ Fig.~\ref{fig:PA_F_Phi}:
\begin{subequations}\label{PA_F_Phi}
\begin{align}
\Phi_{d/m}^{k,k^{\prime},q} &= \sum_{k^{\prime\prime}}\Gamma^{k,k^{\prime\prime}\!,q}_{d/m}\;G_{k^{\prime\prime}}G_{k^{\prime\prime}+q}\,F^{k^{\prime\prime}\!,k^{\prime},q}_{d/m}\;,\\
\Phi_{t/s}^{k,k^{\prime},q}&= \pm\sum_{k^{\prime\prime}}\Gamma^{k,k^{\prime\prime}\!,q}_{t/s}\;G_{k^{\prime\prime}}G_{q-k^{\prime\prime}}F^{k^{\prime\prime}\!,k^{\prime},q}_{t/s}\;.
\end{align}
\end{subequations}

In the Bethe Salpeter Eqs.~(\ref{PA_F_Phi}), we need the one-particle Green's function as an input. For a self-consistent scheme, we hence need to recalculate the one-particle quantities from the two-particle vertices. This
is achieved via the Dyson-Schwinger equation~\eqref{DysSchw}. 

At this stage, it is easy to see that given the fully irreducible vertex $\Lambda_{d/m/s/t}$ as input, the other two-particle vertices and the single-particle self-energy can be self-consistently determined from the parquet equations (\ref{PA_F}), the Bethe-Salpeter (\ref{PA_F_Phi}), the Dyson equation (\ref{GF}) and the Dyson-Schwinger equation (\ref{DysSchw}). If  $\Lambda_{d/m/s/t}$ is known there are as many equations as unknowns; also note that $\Lambda_{s/t}$ can be determined from   $\Lambda_{d/m}$ for $SU(2)$ symmetry.
We solve this set of equations self-consistently using the {\em victory} code~\cite{victory} until convergence is reached. For details of the numerical implementation including how we deal with the high-frequency asymptotics see Ref.~\onlinecite{victory}.

\subsection{Parquet approximation (PA)}
\label{Sec:ParquetApprox}

In the PA~\cite{Bickers-Review} the parquet equations are solved with the lowest order approximation for the fully irreducible vertex $\Lambda_{d/m/s/t}$, i.e., 
\begin{align}\label{parquet_approx}
\Lambda_{d/m}&=\pm U, \nonumber\\ \Lambda_{s}&=2 U,\nonumber\\  \Lambda_{t}&=0.
\end{align}
Although the fully irreducible vertex is frequency and momentum independent, all other vertices are dependent on three Matsubara frequencies and three momenta. 

\subsection{Dynamical vertex approximation}
\label{Sec:ParquetDGA}
In the parquet flavor of dynamical vertex approximation (parquet D$\Gamma$A), the fully irreducible vertex $\Lambda$ is extracted from the impurity problem of a converged dynamical mean-field theory (DMFT) calculation, as described in Ref.~\onlinecite{Valli2015}. That is, the dynamical structure of the fully irreducible vertex is retained (for details see Refs.~\onlinecite{PhysRevB.86.125114,RMPVertex}) but its momentum dependence is neglected, i.e., $\Lambda_{d/m/s/t}^{k,k',q}\equiv\Lambda_{d/m/s/t}^{\nu,\nu',\omega}$. In this work the DMFT impurity problem was solved with the continuous time quantum Monte-Carlo method (CT-QMC) in the hybridization expansion (CT-HYB) scheme as implemented in the software package {\it w2dynamics}\cite{w2dynamics}. The fully irreducible vertex was extracted  from the two-particle Green functions using the inverse parquet equations, following the procedure described in Ref.~\onlinecite{Hoerbinger_MA} (see also Sec.~\ref{Sec_numerical} and Ref.~\onlinecite{RMPVertex}).

\subsection{Eigenvalues  as indicators of possible instabilities}

\label{Sec_eigenvalues}

A second order phase transition with the emergence of a finite order parameter can be identified from the divergence of the susceptibility corresponding to the order parameter. The physical susceptibility $\chi_r(q)$ in a given channel $r=\{d,m,pp\}$ is obtained by summation over fermionic frequencies and momenta of a generalized susceptibility $\chi_r(q)=\sum_{k,k'}\chi^{k,k',q}_r$. The generalized susceptibilities obey Bethe-Salpeter equations analogous to Eq.~\eqref{PA_F_Phi}:
\begin{subequations}\label{BS_susc}
\begin{align}
\chi_{d/m}^{k,k^{\prime},q} &= - \beta G_k G_{k^{\prime}+q}\delta_{kk'} 
+\sum_{k^{\prime\prime}}G_{k}G_{k+q}\;\Gamma^{k,k^{\prime\prime}\!,q}_{d/m}\;\chi^{k^{\prime\prime}\!,k^{\prime},q}_{d/m}\;,\\
%
\chi_{pp,\overline{\up\dn}}^{k,k^{\prime},q} &= - \beta G_k G_{q-k^{\prime}}\delta_{kk'} - \sum_{k^{\prime\prime}}G_{k}G_{q-k}\;\Gamma^{k,k^{\prime\prime}\!,q}_{pp,\overline{\up\dn}}\;\chi^{k^{\prime\prime}\!,k^{\prime},q}_{pp,\overline{\up\dn}}\;,
\end{align}\end{subequations}
with $\Gamma^{k,k^{\prime\prime}\!,q}_{pp,\overline{\up\dn}}=-\Gamma^{k,q-k^{\prime\prime}\!,q}_{pp,{\up\dn}}=\frac{1}{2}[\Gamma^{k,k^{\prime\prime}\!,q}_{t}- \Gamma^{k,k^{\prime\prime}\!,q}_{s}]$.
The instabilities of the above equations can be investigated by solving the following eigenvalue problems instead:
\begin{subequations}
\begin{align}
\sum_{k^{\prime}}\Gamma_{d/m}^{k, k^{\prime},q}\;G_{k^{\prime}}G_{k^{\prime}+q}\; v^{q}(k^{\prime}) = \lambda^q_{d/m}(k) \;v^{q}(k)\;,\\
\sum_{k^{\prime}}\Gamma_{pp,\overline{\up\dn}}^{k, k^{\prime},q}\;G_{k^{\prime}}G_{q-k^{\prime}}\;v^{q}(k^{\prime})=\lambda^q_{pp}(k) \;v^{q}(k)\;.
\label{EV_eq}
\end{align}
\end{subequations} 
Once the eigenvalue $\lambda_r$ in a given channel $r$ approaches $1$, the corresponding susceptibility diverges.  The above equation is diagonal in bosonic frequency $\omega$ and momentum ${\mathbf q}$. The largest eigenvalue typically occurs for $\omega=0$ and a given transfer momentum ${\mathbf q}$, depending on the specific order considered (e.g.\ $r=m$, ${\mathbf q}=(\pi,\pi)$ for N\'eel antiferromagnetic order). Although we do not investigate the system in the ordered phase, we can identify the symmetry of the plausible order parameter by looking at the momentum dependence of the eigenvector $v^{q}(k)$ belonging to the largest (closest to $1$) eigenvalue. The eigenvalue equation in the particle-particle channel for $q=0$ [i.e. ${\mathbf q}=(0,0)$ and $\omega=0$] is thus analogous to the Eliashberg equation for the superconducting gap $\Delta$ 
\begin{equation}
\label{Eliashberg}
-\sum_{k^{\prime}}V(k-k')G_{k^{\prime}}G_{-k^{\prime}}\Delta(k^{\prime})=\Delta(k)\;,
\end{equation}
if we identify $V(k-k')\equiv-\Gamma_{pp,\overline{\up\dn}}^{k, k^{\prime},q=0}=\Gamma^{k,-k^{\prime},q=0}_{pp,{\up\dn}}$.
In our parquet and D$\Gamma$A calculations, however, we retain the full momentum and frequency dependence of the vertex $\Gamma_{pp,{\up\dn}}$, in contrast to the conventional Eliashberg equation.

\subsection{Numerical tools}
\label{Sec_numerical}

The PA and D$\Gamma$A results were obtained using the {\it victory} code \cite{victory}. 
In addition to the published version 1.0 of the  {\it victory} code \cite{victory}, the following operations were implemented for the present work  and will be made available as  {\it victory} version 1.1:
\begin{itemize}
\item Point group symmetry of the square lattice. The main advantage of this step relies in the reduction of virtual memory needed to store the vertices. For a large momentum grid, the memory reduction reaches a factor of 8, but for small clusters used in this work the reduction factor was close to 3  and 4 for the $6\times 6$ and $8\times 8$ cluster, respectively.
Let us note that these cluster sizes are larger than what is doable in quantum Monte Carlo simulations (QMC) or in dynamical cluster approximation (DCA) which (for larger dopings) are restricted to $N_c=16$ sites  (e.g.~$4\times4$ sites) because of the sign problem.\cite{PhysRevLett.115.116402,2018arXiv181010043M}
\item Coarse graining. Since storing the two-particle vertices reaches the limits of random access memory, the number of $\mathbf k$-points is rather restricted. For the Green's function on the other hand, a much finer  $\mathbf k$-grid is possible.  Hence, in the Bethe-Salpeter equations (\ref{PA_F_Phi}) and the Schwinger-Dyson equation
(\ref{DysSchw}) for calculating the self-energy, we use the Green's functions 
$$
G_{\widetilde{\mathbf{k}},\nu}=\frac{1}{i\nu+\mu-\epsilon_{\widetilde{\mathbf{k}}} -\Sigma_{\mathbf{k},\nu}},
$$
on a much finer $\widetilde{\mathbf{k}}$-grid than the coarse $\mathbf{k}$-grid for which we know the self-energy or vertex. The number of $\widetilde{\mathbf{k}}$-points was typically 100 times larger than the number of cluster momenta $\mathbf{k}$ and the  self-energy was taken as constant over a 2-dimensional patch
 surrounding a given $\mathbf{k}$ point.
This has the advantage that finer bandstructure $\epsilon_{\widetilde{\mathbf{k}}}$ effects can be resolved. The coarse graining approach also significantly improves convergence of the parquet loop.

\end{itemize}

\begin{figure}[htp]
\centering
\includegraphics[width=0.49\textwidth]{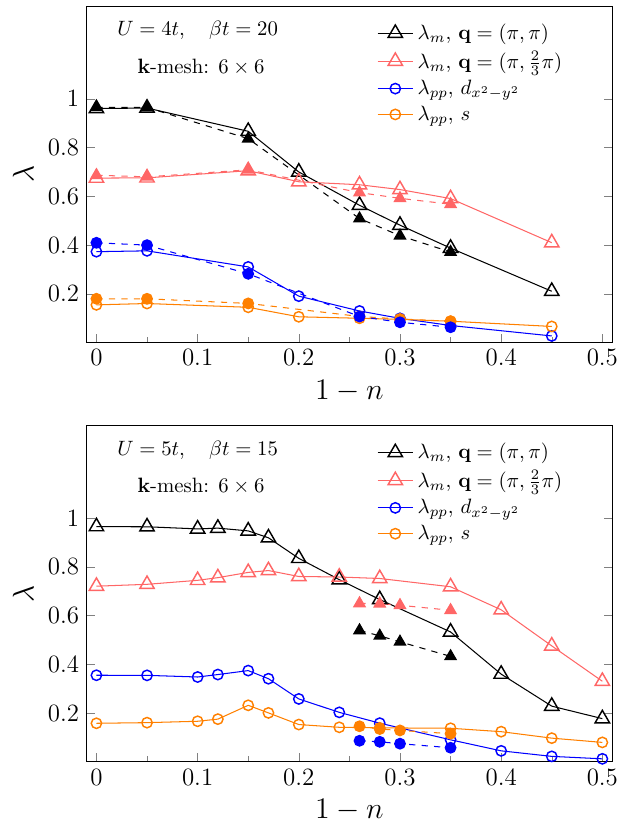}
\caption{Dominant eigenvalues in the magnetic ($\lambda_m$, triangles) and particle-particle ($\lambda_{pp}$, circles) channels as a function of doping for two different values of the interaction: $U=4t$ (top) and $U=5t$ (bottom). Different triangle colors (red and black) denote different transfer momenta ${\bf q}$. Different circle colors (blue and orange) denote different symmetries of the corresponding eigenvectors ($d_{x^2-y^2}$-wave and triplet $s$-wave, respectively) at transfer momentum ${\bf q}=(0,0)$. The empty symbols are obtained within the PA, the full symbols are from D$\Gamma$A. The momentum cluster size is $6\times 6$ and the temperature is $\beta t =20$ for $U=4t$ and $\beta t =15$ for $U=5t$.}
\label{Fig_lambda_of_n}
\end{figure}


\section{Results}

\label{Sec:Results}

\subsection{Dominant eigenvalues as a function of doping} 
\label{Sec:DominantEW}

Let us start by noting, that the calculations presented in this section are all in the paramagnetic phase.  No direct evidence of a magnetic or superconducting phase transition through the divergence of the susceptibility has been found for the parameters investigated, because of the high temperature. However, the analysis of the eigenvalues in different channels and for different transfer momenta $\mathbf{q}$  gives us an indication of a strong interplay between the magnetic and particle-particle channels, and of the possible ordering at lower temperatures. Figure~\ref{Fig_lambda_of_n} shows the dominant eigenvalues  for $\beta t \equiv t/ T=20$, $U=4t$ (top) and $\beta t =15$, $U=5t$ (bottom) as a function of doping $(1-n)$. The calculations were done for the $6\times6$ momentum cluster and two different approximations: PA and parquet D$\Gamma$A.\cite{footnote3} Since the convergence of the D$\Gamma$A computations is significantly worse than PA computations, we show the D$\Gamma$A results only for selected points in the parameter space. 

\begin{figure}[!htb]
\centering
\includegraphics[width=0.49\textwidth]{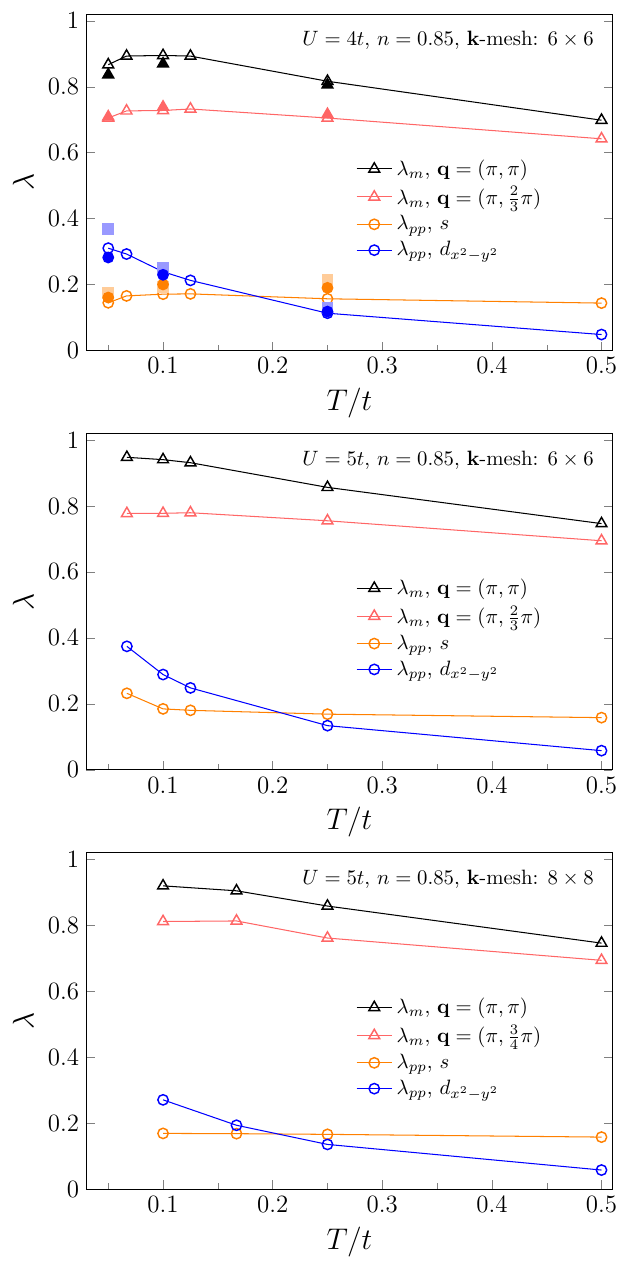}
\caption{Temperature dependence of the dominant eigenvalues in the magnetic ($\lambda_m$, triangles) and particle-particle ($\lambda_{pp}$, circles) channels for $n=0.85$ and different values of the interaction: $U=4t$ (top) and $U=5t$ (middle and bottom), and different cluster sizes: $6\times 6$ (top and middle) vs. $8\times 8$ (bottom).
 Color coding is identical to Fig.~\ref{Fig_lambda_of_n}; empty symbols are obtained within the PA, full within the D$\Gamma$A; square symbols are ladder D$\Gamma$A results discussed later in Section~\ref{sec:ladder}. }
\label{Fig_lambda_T}


\end{figure}

\subsubsection{Small dopings}
\label{Sec:Smalldoping}
Close to half-filling we find that the overall dominant eigenvalue is the  $\mathbf{q}=(\pi,\pi)$ magnetic eigenvalue, which indicates very strong antiferromagnetic fluctuations. The  dominant eigenvalue in the particle-particle channel $\lambda_{pp}$ is found at $\mathbf{q}=(0,0)$. Looking at the fermionic momentum dependence of the eigenvectors belonging to $\lambda_{pp}$, we can identify the symmetry of the related superconducting order parameter (see Sec.~\ref{Sec_eigenvalues}). For dopings up to $\sim0.2$ the dominant symmetry is spin-singlet, $d_{x^2-y^2}$-wave (blue circles in Fig.~\ref{Fig_lambda_of_n}) both for $U=4t$ and $U=5t$. The $d_{x^2-y^2}$-wave eigenvalue follows very closely the changes of the antiferromagnetic eigenvalue with doping. We observe a slightly different behavior for the higher value of the interaction, $U=5t$, where there is a slight dome (broad maximum) in the $pp$-eigenvalue doping dependence around $n=0.85$. This dome is not present in the $\mathbf{q}=(\pi,\pi)$ magnetic eigenvalue curve (black triangles), but it is present (slightly shifted) in the $\mathbf{q}=(\pi,\frac{2}{3}\pi)$ magnetic eigenvalue curve (red triangles). 
As we will see below, the triplet $s$-wave channel is  connected to the fact that the  antiferromagnetic fluctuations become incommensurate as well as to the frequency dependence of the vertex. Please note, that due to the small size of the momentum cluster ($6\times 6$), we cannot study the effects of slight incommensurability, but when it occurs, the incommensurability is already quite large, i.e. $\mathbf{q}=(\pi,\frac{2}{3}\pi)$.


For $U=5t$ it was not possible to converge the D$\Gamma$A computations for dopings smaller than $n=0.74$. In contrast to the case of $U=4t$, where the inclusion of dynamical fully irreducible vertex $\Lambda$ had relatively little effect on both the eigenvalues and convergence properties (it took much longer to converge the D$\Gamma$A calculations, but it was possible for all dopings up to $\beta t=20$), for $U=5t$ we could not obtain convergence even for much higher temperatures. It is likely that for higher values of the interaction, the diagrams included in the fully irreducible vertex $\Lambda$ become more important and  the impurity problem used for generating the dynamical $\Lambda$ needs to be self-consistently updated. This is not done in the present computational scheme.


\subsubsection{Larger dopings}
\label{Sec:Largedoping}
For larger dopings the magnetic $\mathbf{q}=(\pi,\pi)$ eigenvalue becomes smaller and another eigenvalue dominates: $\lambda_{m}$ at $\mathbf{q}=(\pi,\pi-\delta)$ (red triangles  in Fig.~\ref{Fig_lambda_of_n}). That is, the magnetic susceptibility has a peak at $\mathbf{q}=(\pi,\pi-\delta)$. Since we can only resolve relatively few $\mathbf{q}$ points, the value of $\delta$ depends on the cluster size used (for the \mbox{$6\times6$} cluster, $\mathbf{q}=(\pi,\frac{2}{3}\pi)$). In an infinite system we can expect incommensurate magnetic fluctuations, as seen in Refs.~\onlinecite{Schulz1990,Freericks1995,Jarrell1997,Yamase2016,Schaefer2016,Kozik_PRB_2017}. 

For both values of the interaction shown, once the magnetic $\mathbf{q}=(\pi,\pi)$ eigenvalue starts to steeply decrease with doping, the  $d_{x^2-y^2}$-wave eigenvalue follows suit. The incommensurate magnetic fluctuations remain however strong in a larger range of doping,  indicated by a large  $\mathbf{q}=(\pi,\frac{2}{3}\pi)$ eigenvalue, which starts decreasing only at around $n=0.65$. The $d_{x^2-y^2}$-wave symmetry of the $pp$-eigenvector is evidently strongly related to antiferromagnetism, and incommensurate magnetic fluctuations do not prevent the $d_{x^2-y^2}$-wave eigenvalue from decreasing with doping. At large dopings, another symmetry of the $pp$ eigenvector wins, i.e.,  odd-frequency triplet $s$-wave (orange circles in Fig.~\ref{Fig_lambda_of_n}). The eigenvalues are however small at this temperature and it is not possible to predict on that basis that triplet $s$-wave superconductivity could occur at low temperature in this doping regime.

\begin{figure*}
\centering
\begin{minipage}{\textwidth}
\includegraphics[width=\textwidth]{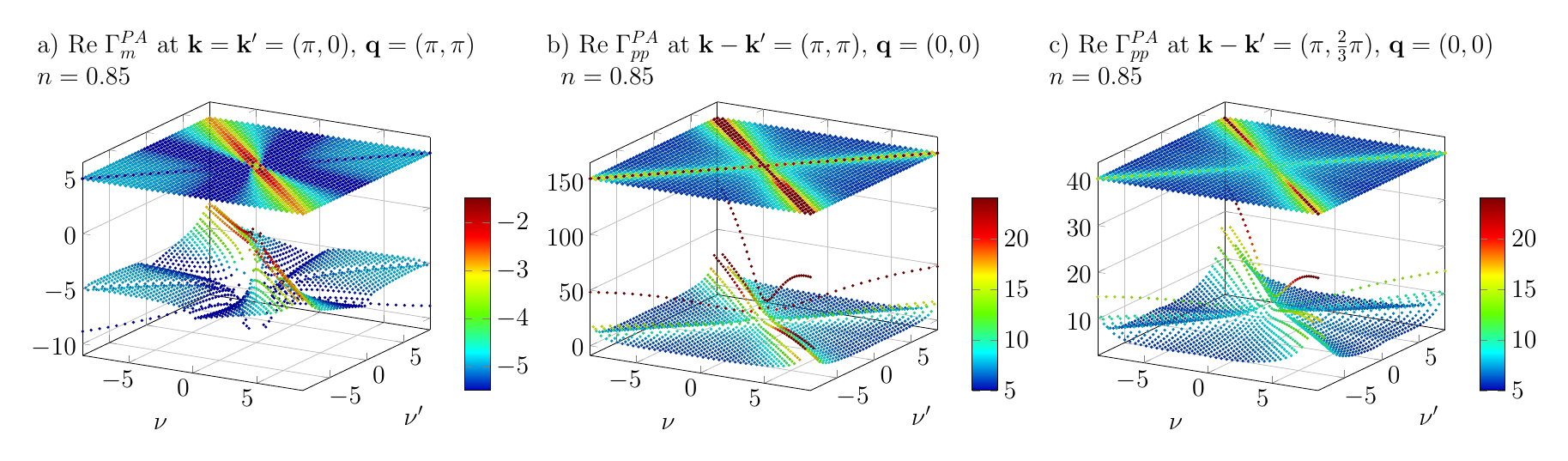}
\caption{Fermionic frequency dependence of the real part of the irreducible vertex in PA for  $n=0.85$, $U=5t$, $\beta t =15$, $6\times 6$ cluster. Left: magnetic channel, $\mathrm{Re}\;\Gamma_m^{\nu,\nu',\omega=0}$. Middle and right:  particle-particle channel, $\mathrm{Re}\;\Gamma_{pp,{\up\dn}}^{\nu,\nu',\omega=0}$, for $\mathbf q=0$ and two different (${\mathbf{k}-\mathbf{k}'}$)-momenta as denoted in the figure. }  
\label{Fig_gamma_n085}

\end{minipage}

\end{figure*}

\subsubsection{The effect of dynamical $\Lambda$} 
\label{Sec:VertexDynamics}
For the smaller value of the interaction, $U=4t$ shown in the upper panel of Fig.~\ref{Fig_lambda_of_n}, including the dynamics of the fully irreducible vertex $\Lambda$ (in parquet D$\Gamma$A) does not have a qualitative influence on the behavior of the eigenvalues with doping. The D$\Gamma$A results (full symbols) lie almost on top of the PA values (empty symbols) for small dopings, and the magnetic and $d_{x^2-y^2}$-wave eigenvalues  are slightly suppressed in D$\Gamma$A as compared to PA as the doping is increased. That is, the vertex dynamics  reduces antiferromagnetic fluctuations, which further reduces the $d_{x^2-y^2}$-wave eigenvalue, cf.~Ref.~\onlinecite{Kitatani2018}.

The suppression of magnetic fluctuations through the dynamics of the fully irreducible vertex is significantly stronger for the larger interaction, $U=5t$ shown in lower panel of Fig.~\ref{Fig_lambda_of_n}. Here we were able to obtain convergent results only for larger dopings, but again the stronger suppression of antiferromagnetic fluctuations in D$\Gamma$A leads to proportionally stronger suppression of the $d_{x^2-y^2}$-wave eigenvalue. Since the triplet $s$-wave eigenvalue is not influenced and remains almost the same in PA and  D$\Gamma$A, there is a qualitative difference between the two approximations: for a small range of dopings the dominant $pp$ eigenvalue is of singlet  $d_{x^2-y^2}$-wave symmetry in PA whereas it is of triplet $s$-wave symmetry in  D$\Gamma$A. It would be interesting to see how this difference develops while the temperature is lowered and whether the triplet $s$-wave pairing could be the dominant order at low temperature. The currently feasible frequency box sizes and momentum clusters do not allow for calculations at lower temperatures in the present implementation of the {\it victory} code.

\begin{figure*}
\centering
\begin{minipage}{\textwidth}
\includegraphics[width=\textwidth]{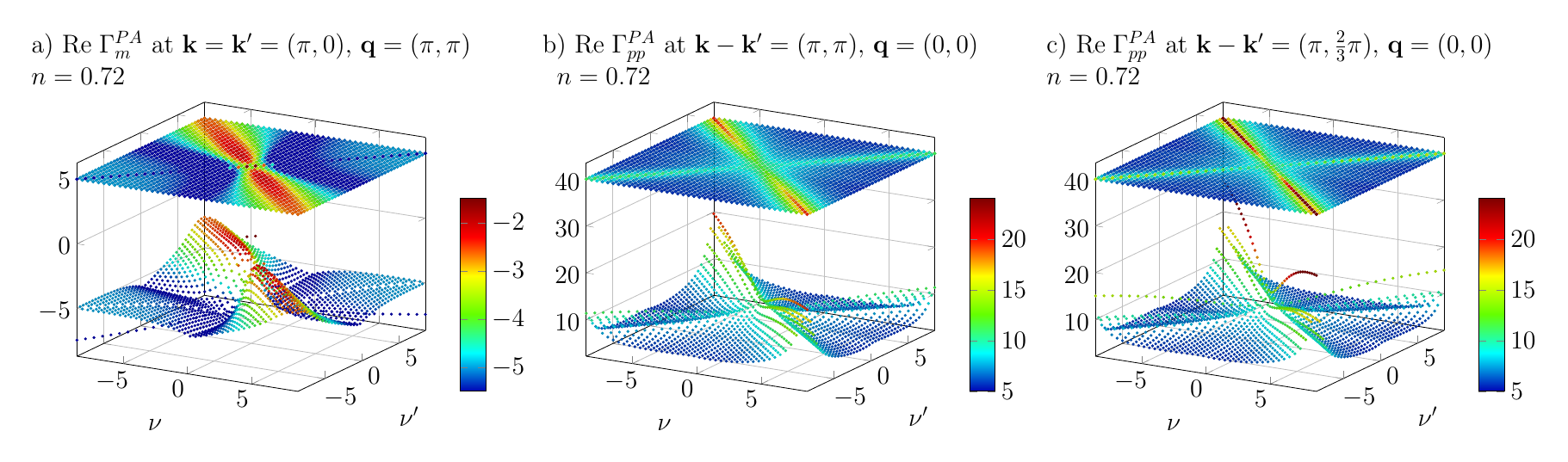}
\includegraphics[width=\textwidth]{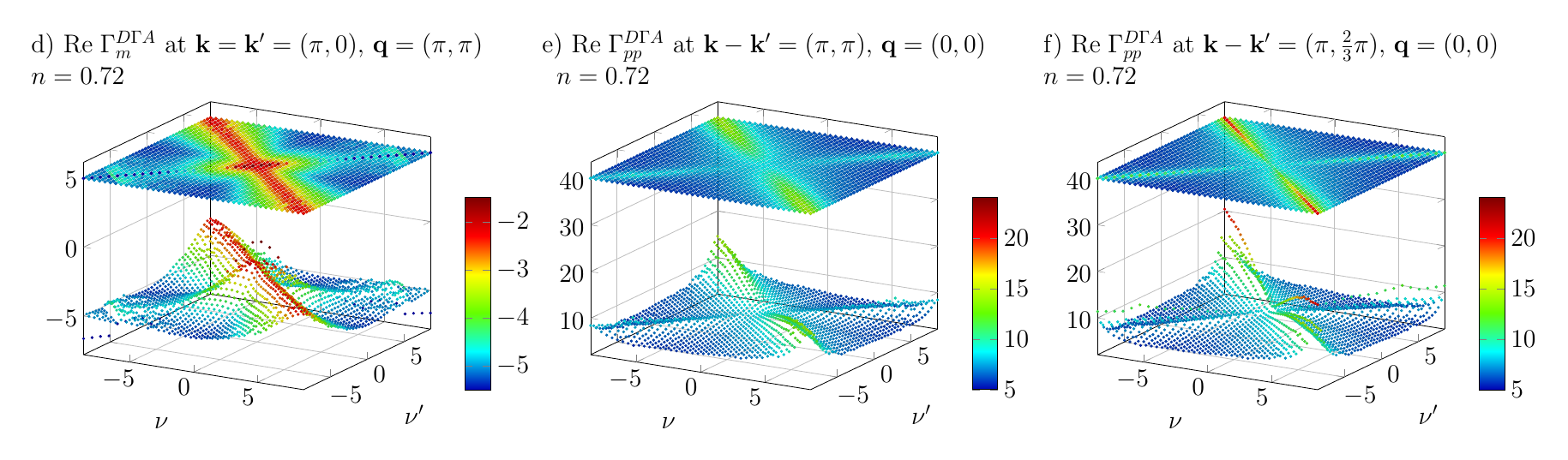}
\includegraphics[width=\textwidth]{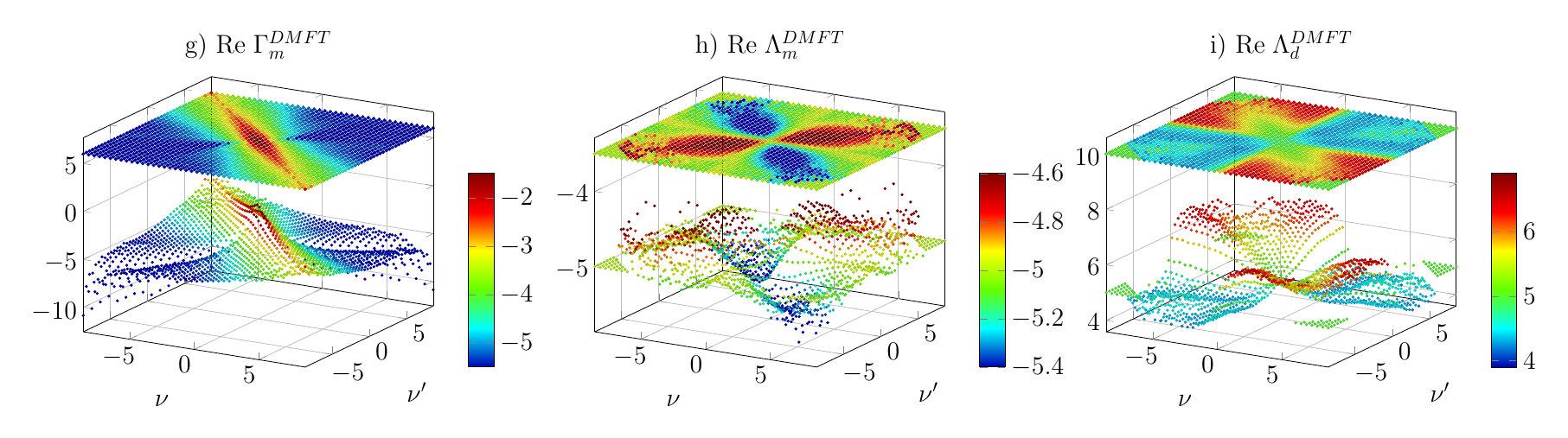}
\caption{Fermionic frequency dependence of the real part of the irreducible vertex in PA (top row) and D$\Gamma$A (middle row) for  $n=0.72$, $U=5t$, $\beta t=15$, $6\times 6$ cluster. Left: magnetic channel, $\mathrm{Re}\;\Gamma_m^{\nu,\nu',\omega=0}$. Middle and right:  particle-particle channel, $\mathrm{Re}\;\Gamma_{pp,{\up\dn}}^{\nu,\nu',\omega=0}$, for $\mathbf q=0$ and two different (${\mathbf{k}-\mathbf{k}'}$)-momenta as denoted in the figure. The last row shows the local DMFT magnetic vertex $\Gamma_m^{DMFT}$ (left) and   the local fully irreducible DMFT vertex $\Lambda_r$ in the magnetic ($r=m$, central column) and density channel  ($r=d$, right) at $\omega=0$. }
\label{Fig_gamma_n072}

\end{minipage}

\end{figure*}

\subsection{Temperature dependence of the eigenvalues} 
\label{Sec:Tdep}
In order to identify possible phase transitions we look at the temperature dependence of the eigenvalues. We are computationally limited to still relatively high temperatures $T\gtrsim0.06t$ ($\beta t \lesssim 15$) for $U=5t$ and $T\gtrsim0.05t$ ($\beta t \lesssim 20)$ for $U=4t$. Lower temperatures require a larger frequency box and/or larger momentum clusters, for a proper convergence of the parquet equations and stable results with respect to the size of the frequency box.  We have observed that increasing the momentum cluster significantly improves convergence of the parquet equations, and a smaller frequency box can be used. The $8\times8$ cluster calculations presented in this paper were obtained with up to $64$ frequencies. For the  $6\times6$ cluster, however,  we were able to go to as many as $160$ (positive and negative) frequencies. Numerical convergence  with respect to frequency box size is presented  in Appendix \ref{Apppendix:convergence}. 

In Fig.~\ref{Fig_lambda_T} we present the temperature dependence of the four eigenvalues of interest for the $6\times6$ cluster (top panel: $U=4t$, middle panel: $U=5t$) and $8\times8$ cluster (bottom panel, $U=5t$), at the filling of $n=0.85$ (close to optimal for the $d_{x^2-y^2}$-wave order). The qualitative behavior with temperature is similar in all three cases.

For the smaller interaction, $U=4t$,  the eigenvalues are overall smaller than for $U=5t$. Although the magnetic eigenvalues (triangles in Fig.~\ref{Fig_lambda_T}) are dominant in the entire temperature regime and generally grow with lowering $T$, we see that their temperature dependence flattens and for $U=4t$ there is even a small decrease upon lowering $T$ visible. On the contrary, the $d_{x^2-y^2}$-wave eigenvalue (blue circles in Fig.~\ref{Fig_lambda_T})  increases with lowering the temperature, more steeply for stronger interaction. It is clearly visible, that although for high temperatures the $s$-wave eigenvalue (orange circles in Fig.~\ref{Fig_lambda_T}) is the bigger one, upon lowering $T$, the $d_{x^2-y^2}$-wave symmetry wins. It is not possible to predict the transition temperature though from the data available. Actually, one might conjecture that the parquet equations respect the Mermin-Wagner theorem\cite{Mermin1966,Bickers1992} with no phase transition at finite $T$. A numerical verification of this conjecture would however require an analysis at lower temperatures than presently possible.

The comparison of the temperature dependence of the eigenvalues for the same parameters ($U=5t$, $n=0.85$) and two different momentum cluster sizes is shown  in the middle panel ($6\times6$) and bottom panel ($8\times8$) of Fig.~\ref{Fig_lambda_T}). The results qualitatively agree (note, that the red triangles are for different ${\bf q}$s in the two plots). Since calculations for a smaller cluster size do not converge and for a larger cluster size are not feasible, the assessment of the cluster size effect is very limited and therefore we cannot directly predict the infinite cluster limit. 

For the two highest temperatures shown in Fig.~\ref{Fig_lambda_T} in the $pp$ channel also other eigenvalues, corresponding to singlet $s$-wave, $p$-wave and $d_{xy}$-wave symmetries, are of similar magnitude. They however remain small (or become even smaller) upon lowering the temperature and we do not show them in Fig.~\ref{Fig_lambda_T}.


\begin{figure*}
\centering
\begin{minipage}{\textwidth}
\includegraphics[width=1.02\textwidth]{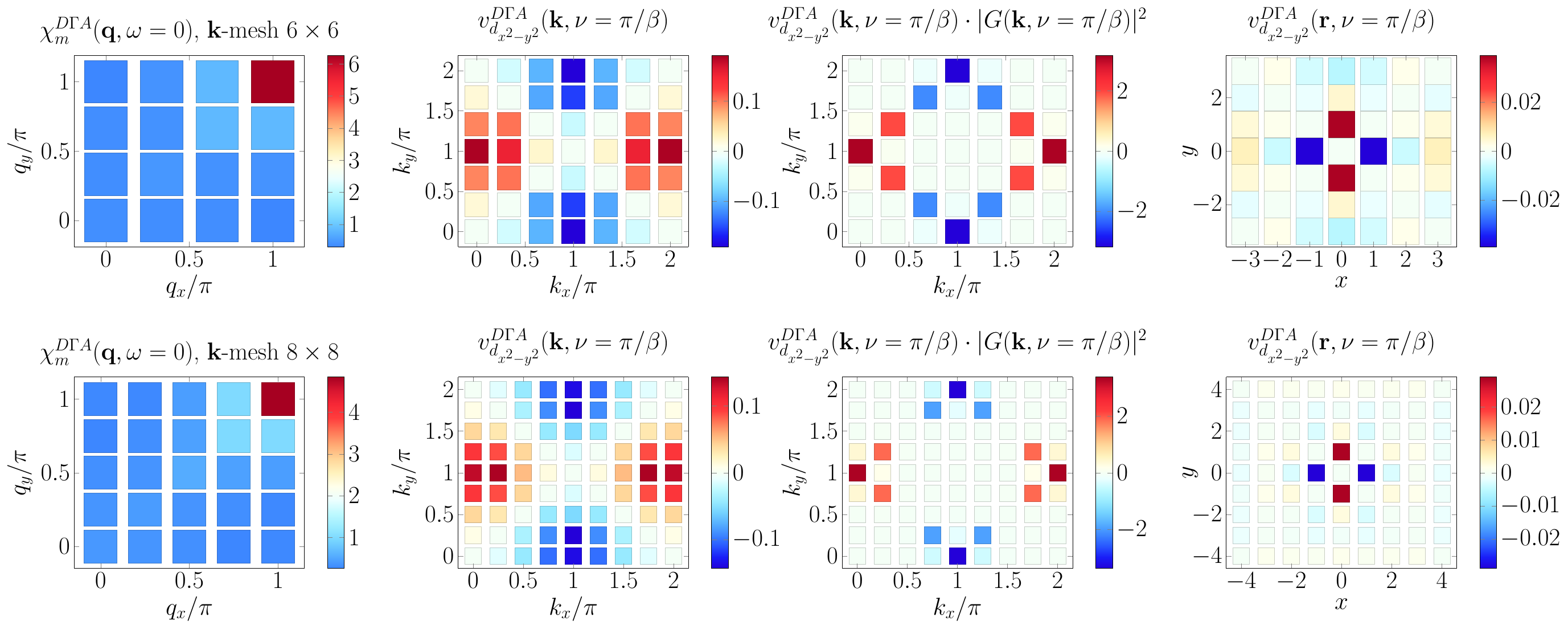}
\caption{Left: Magnetic susceptibility $\chi_m({\bf q}, \omega=0)$ vs.~$q_x$ and $q_y$ as obtained from D$\Gamma$A at $U=4t$, $\beta t=20$, $n=0.85$ for two different cluster sizes: $6\times6$ (upper row) and $8\times8$ (lower row). Left-middle column: Eigenvector $v_{d_{x^2-y^2}}({\bf k}, \nu=\pi/\beta)$ corresponding to the dominant eigenvalue in the particle-particle channel. Right-middle column: Eigenvector $v_{d_{x^2-y^2}}({\bf k}, \nu=\pi/\beta)$ from the left-middle column projected onto the Fermi surface (i.e. multiplied by $|G|^2$ at the lowest Matsubara frequency). Right:
Fourier transform $v_{d_{x^2-y^2}}({\bf r}, \nu=\pi/\beta)$ of the eigenvector from the left-middle column. The unit of distance is the  lattice constant $a=1$.}
\label{Fig_susc_n085}

\end{minipage}

\end{figure*}

\subsection{Dynamical vertices}
\label{Sec:DynVertex}
The parquet equations \eqref{PA_F} mix the  different channels and generate a complex frequency dependence, even in the PA with a static $\Lambda_r=\pm U$ as a starting point. We focus here, as in the eigenvalue analysis, on the interplay between the magnetic and particle-particle channel. We present the frequency dependence of the irreducible vertex in the magnetic and $pp$ channel for two different, representative dopings: $n=0.85$ [close to optimal for $d_{x^2-y^2}$-wave symmetry with dominant antiferromagnetic fluctuations] and $n=0.72$ [where magnetism with ordering vector  $\mathbf{q} =(\pi,\frac{2}{3}\pi)$ starts to dominate] for the $6\times 6$ cluster, $U=5t$ and $\beta t = 15$.

\subsubsection{Filling $n=0.85$}

In Fig.~\ref{Fig_gamma_n085} we show, for the PA at filling $n=0.85$, the fermionic frequency dependence of the real part of the irreducible vertex  in (a) the magnetic channel,  $\mathrm{Re}\;\Gamma_m^{\nu,\nu',\omega=0}$  for $\mathbf{k}=\mathbf{k}'=(\pi,0)$ and $\mathbf{q}=(\pi,\pi)$, (b-c) the  particle-particle  channel, $\mathrm{Re}\;\Gamma_{pp, {\up\dn}}^{\nu,\nu',\omega=0}$ for $\mathbf{q}=(0,0)$ and two different combinations of fermionic momenta: (b) $\mathbf{k}-\mathbf{k}'=(\pi,\pi)$  and (c) $\mathbf{k}-\mathbf{k}'=(\pi,\frac{2}{3}\pi)$. The momenta have been set so that the vertices are maximal: $\Gamma_m$ is maximal at transfer momentum $\mathbf{q}=(\pi,\pi)$ since at $n=0.85$ commensurate antiferromagnetic fluctuations dominate, cf.~Fig.~\ref{Fig_lambda_of_n});  whereas $\Gamma_{pp}$ is maximal at transfer momentum $\mathbf{q}=(0,0)$. We can see not only a strong frequency dependence with a characteristic diagonal structure~\cite{PhysRevB.86.125114} in all vertices shown, but also a strong fermionic momentum dependence of the ${pp}$ vertex. Through the parquet-equations \eqref{PA_F} a peak in $\Gamma_m$ at momentum $\mathbf{q}=(\pi,\pi)$ (Fig.~\ref{Fig_gamma_n085} a) is transferred to a peak in $\Gamma_{pp}$ at $\mathbf{q}=(0,0)$ and $\mathbf{k}-\mathbf{k}'=(\pi,\pi)$ (Fig.~\ref{Fig_gamma_n085} b). 
Such momentum dependence of $\Gamma_{pp}$ in the normal state, above $T_c$ supports the emergence of a superconducting order parameter of the $d_{x^2-y^2}$-wave symmetry when the temperature is lowered below $T_c$. For $\mathbf{k}-\mathbf{k}'=(\pi,\frac{2}{3}\pi)$  (Fig.~\ref{Fig_gamma_n085} c) the $\Gamma_{pp}$  vertex is significantly smaller and the frequency dependent features are less pronounced, at least for this doping close to half filling. We also observe a strong suppression of the $pp$ vertex for small fermionic frequencies, in agreement with Ref.~\onlinecite{Kitatani2018}.

\subsubsection{Filling $n=0.72$}

Fig.~\ref{Fig_gamma_n072}(a-c) shows, analogously to Fig.~\ref{Fig_gamma_n085},  the fermionic frequency dependence of the real part of the irreducible vertices  in the magnetic and $pp$ channels from the PA, but now for the filling of $n=0.72$, comparing PA, D$\Gamma$A and the local DMFT vertices. As already indicated by the eigenvalues, for this filling the magnetic vertex is slightly smaller at $\mathbf{q}=(\pi,\pi)$ than for $n=0.85$ and as a result, $\Gamma_{pp}$ at $\mathbf{q}=(0,0)$ and $\mathbf{k}-\mathbf{k}'=(\pi,\pi)$ is significantly smaller. Note that tiny differences in the eigenvalues that are close to one reflect large differences in the respective vertices themselves. In case of $n=0.72$, it is the $\mathbf{k}-\mathbf{k}'=(\pi,\frac{2}{3}\pi)$ value of $\Gamma_{pp}$ which is larger (Fig.~\ref{Fig_gamma_n072}c). 

In Fig.~\ref{Fig_gamma_n072}(d-f) the respective D$\Gamma$A results for the dynamical structure of the vertices are shown. The overall structure is identical to the PA. The magnitude of $\Gamma_{m}$ at $\mathbf{q}=(\pi,\pi)$ and $\Gamma_{pp}$ at $\mathbf{q}=(0,0)$ and $\mathbf{k}-\mathbf{k}'=(\pi,\pi)$ is however  smaller; particularly in the case of $\Gamma_{pp}$. This is also reflected in the comparison between PA and D$\Gamma$A eigenvalues shown in Fig.~\ref{Fig_lambda_of_n}. The $pp$ vertex $\Gamma_{pp}$ at $\mathbf{q}=(0,0)$ and $\mathbf{k}-\mathbf{k}'=(\pi,\frac{2}{3}\pi)$ is not much smaller in D$\Gamma$A than in PA.
 

In Fig.~\ref{Fig_gamma_n072}(h-i) the real part of the DMFT fully irreducible vertex $\Lambda_{m/d}^{\nu,\nu',\omega=0}$ used as input to the parquet equations in D$\Gamma$A is shown. For comparison, also the local DMFT vertex $\Gamma_m^{\nu,\nu',\omega=0}$ is shown (Fig.~\ref{Fig_gamma_n072} g).

\begin{figure*}
\centering
\begin{minipage}{\textwidth}
\includegraphics[width=\textwidth]{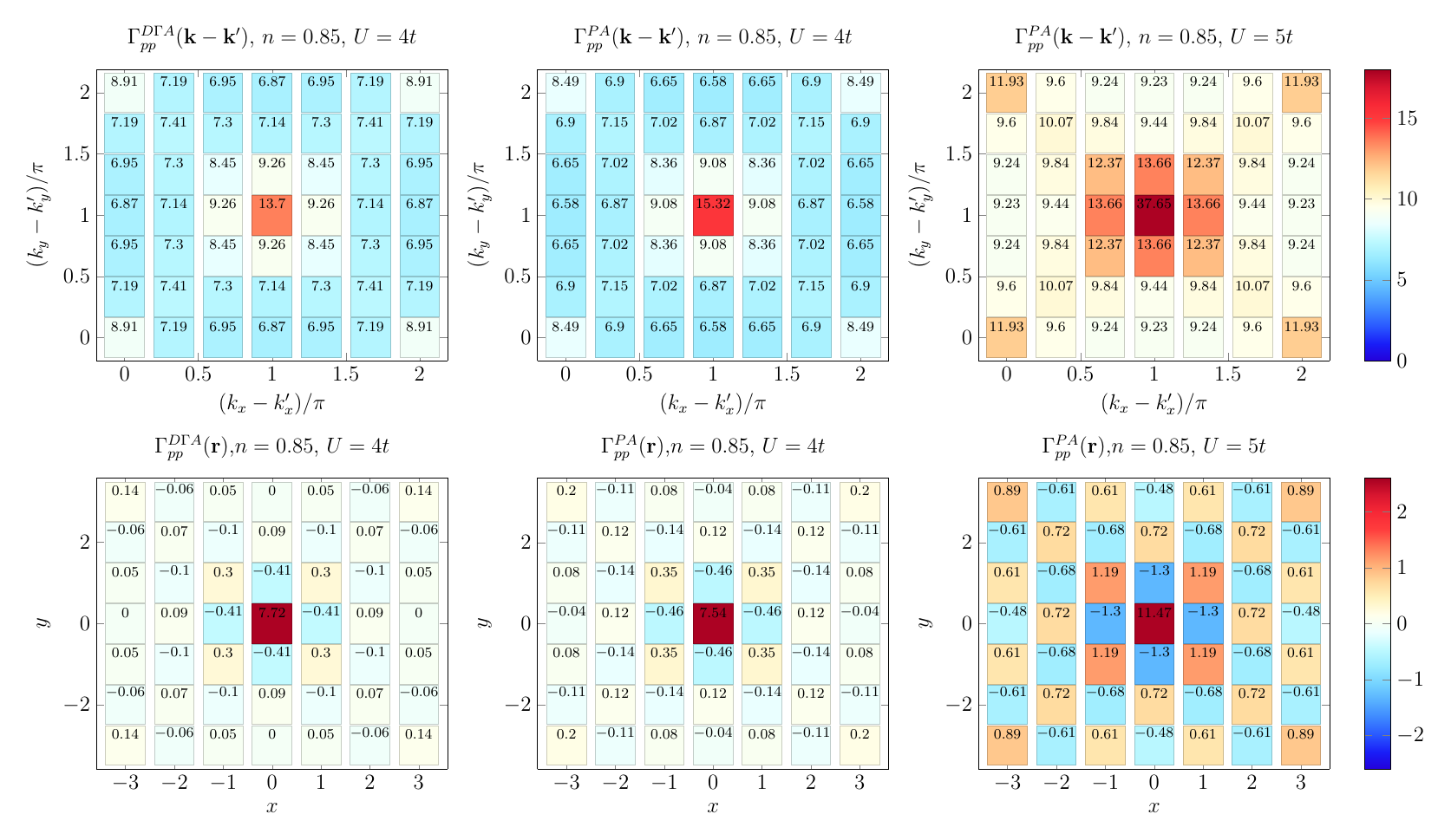}
\caption{Particle-particle vertex $\Gamma_{pp\uparrow\downarrow}$ for $n=0.85$  obtained on a $6\times 6$ momentum grid for $U=4t$ and $\beta t=20$ from  D$\Gamma$A (left column) and PA (middle column) calculations. The $\Gamma_{pp\uparrow\downarrow}$ in the right column is from a PA calculation for a higher value of $U=5t$ and  slightly higher temperature $\beta t=15$. Upper row: Vertex in momentum space $\Gamma_{pp, {\up\dn}}({\bf k}-{\bf k}')$ for $\omega=0$, ${\bf q}=(0,0)$ and $\nu=-\nu'=\pi/\beta$. Lower row: Fourier transform of $\Gamma_{pp, {\up\dn}}({\bf k}-{\bf k}')$ to real space ($x$,$y$).}  
\label{Fig_vect_n085}
\end{minipage}

\end{figure*}

\subsection{Eigenvectors:  $d_{x^2-y^2}$-wave pairing}
\label{dwave}

As already mentioned in Sec.~\ref{Sec_eigenvalues} the analysis of the symmetries of the eigenvectors corresponding to the largest eigenvalues allows us to predict the symmetry of the order parameter once the eigenvalue  as a function of temperature crosses one and the respective susceptibility diverges. In the following, we hence study the symmetry of the eigenvector that corresponds to the largest eigenvalue in Eq.~(\ref{EV_eq}), which presumably diverges at lower temperatures. We start the analysis by presenting the  D$\Gamma$A results for the filling of $n=0.85$. As expected for this doping~\cite{bulut,Maier_2006}, here the symmetry of the dominant eigenvector is---as we will see---$d_{x^2-y^2}$-wave. The results obtained in the PA are qualitatively the same.

In Fig.~\ref{Fig_susc_n085} the momentum dependence of the dominant eigenvector $v_{d_{x^2-y^2}}$ for the first Matsubara frequency is shown (left-middle column) for the two cluster sizes  $6\times6$ and $8\times8$ at $n=0.85$, $U=4t$. For both cluster sizes the pattern is the same: we observe a sign change of the eigenvector (the predicted order parameter) as we move along the Fermi surface ($v_{d_{x^2-y^2}}$ projected on the Fermi surface is shown in the right-middle column) with nodes at $k_{x}=\pm k_y$ visible for the $8\times8$ cluster. For the $6\times6$ cluster we see only the sign change, as there are no $k_{x}=\pm k_y$ points that would lie on the Fermi surface for this grid. 
The sign change of $v({\bf k})$ together with the fact that it is from the singlet $pp$ channel shows that this eigenvalue is of ${d_{x^2-y^2}}$-symmetry (the dominant contribution to $v({\bf k})$ is proportional to $\cos(k_x)-\cos(k_y)$). Hence, we have already labeled it correspondingly in Fig.~\ref{Fig_susc_n085}. 

In the right column of Fig.~\ref{Fig_susc_n085}, the Fourier transform of $v({\bf k})$ from the left-middle column into real space is shown, i.e. $v({\bf r})$, where ${\bf r}$ measures the distance to surrounding sites in units of the lattice constant. We observe that the eigenvector $v$ is strongly peaked at the positions of nearest neighbors and changes sign depending on direction, which is in accordance with expectations for the  ${d_{x^2-y^2}}$-symmetry.

\subsubsection{Relation to magnetic susceptibility}

In the left column of Fig.~\ref{Fig_susc_n085}, the $\mathbf{q}$-dependence of the corresponding magnetic susceptibility $\chi_m(\mathbf{q},\omega=0)$ is shown. It is peaked at $\mathbf{q}=(\pi,\pi)$. This magnetic ordering vector connects those points on the Fermi surface, where the ${pp}$ eigenvector in the middle  column of  Fig.~\ref{Fig_susc_n085} is largest and has opposite sign, i.e.,  $\mathbf{k}=(0,\pi)$ and  $\mathbf{k'}=(\pi,0)$ and (cubic-)symmetrically related ones. It is the  $\mathbf{q}=(\pi,\pi)$ that to a large extent determines the sign pattern in momentum of the superconducting order parameter. This is seen from the Eliashberg Eq.~\eqref{Eliashberg}: For repulsive interaction ($V(k-k')>0$), there has to be a sign change of the order parameter $\Delta(k)$ between the points $k$ and $k'$ connected by  $\mathbf{q}=(\pi,\pi)$. Please recall that our effective interaction, or pairing glue, is given by the irreducible vertex in the particle-particle channel,  $V(k-k')\equiv-\Gamma_{pp,\overline{\up\dn}}^{k, k^{\prime},q=0}$. This in turn, is essentially given by the magnetic susceptibility at  ${\mathbf q}={\mathbf k}-{\mathbf k}'$ since it is the largest particle-particle irreducible contribution.

\begin{figure*}
\centering
\begin{minipage}{\textwidth}
\includegraphics[width=1.02\textwidth]{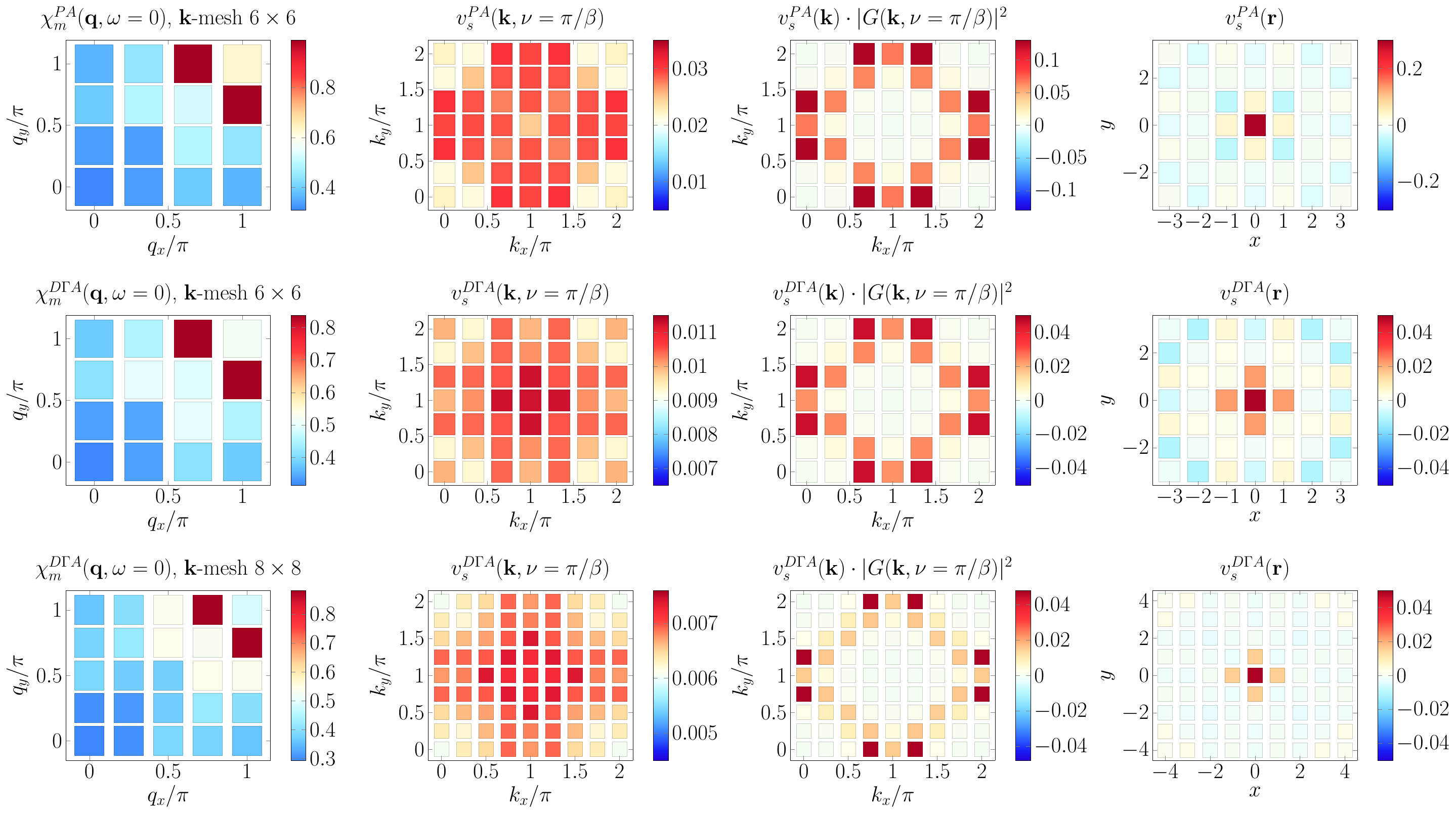}
\caption{Left column: Magnetic susceptibility $\chi_m({\bf q}, \omega=0)$ vs $q_x$, $q_y$, as obtained from PA (upper row) and D$\Gamma$A (middle and lower row) at $U=5t$, $n=0.72$ for a  $6\times6$ (upper and middle row) and $8\times8$ cluster (lower row). Left-middle column: Eigenvector $v_s({\bf k}, \nu=\pi/\beta)$ corresponding to the dominant eigenvalue in the particle-particle channel. Right-middle column: Eigenvector $v_s({\bf k}, \nu=\pi/\beta)$ from the left-middle column projected onto the Fermi surface (i.e. multiplied by $|G|^2$ at the lowest Matsubara frequency). Right column: Fourier transform $v_s({\bf r}, \nu=\pi/\beta)$ of the eigenvector from the left-middle column. }
\label{Fig_susc_n072}

\end{minipage}

\end{figure*}

\begin{figure}
\centering
\includegraphics[width=0.45\textwidth]{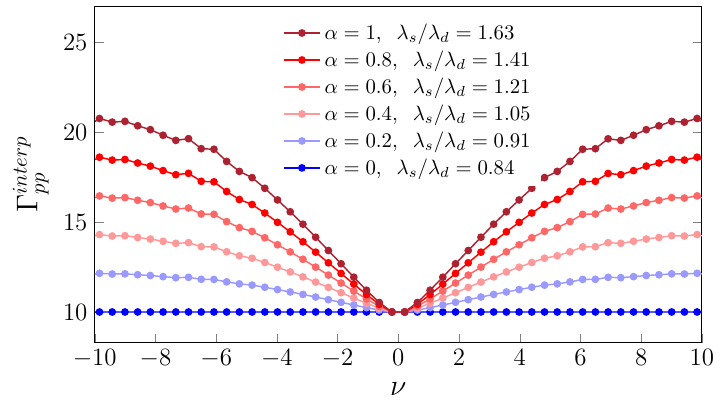}
\caption{Fermionic frequency $\nu$ dependence of the interpolative $pp$ vertex $\Gamma_{pp,{\up\dn}}^{interp}$
taken in the calculation of the eigenvalues in the pairing channel as explained in the text. In the figure only the antidiagonal ($\nu=-\nu'$) of the $\Gamma_{pp,{\up\dn}}^{interp}$ is shown for $\omega=0$ and ${\mathbf{k}-\mathbf{k}'}=(\pi,\frac{2}{3}\pi)$, $\mathbf{q}=(0,0)$. The topmost curve (maroon; $\alpha=1$) is the $\nu=-\nu'$ 2D cut of the 3D D$\Gamma$A  vertex in  Fig.~\ref{Fig_gamma_n072}f) at $n=0.72$, $U=5t$, $\beta t=15$, on a $6\times 6$ cluster. In the legend the ratio of the triplet $s$-wave (odd-frequency) eigenvalue to the $d_{x^2-y^2}$-wave eigenvalue $\lambda_s/\lambda_d$ for the respective value of $\alpha$ is also given. The bottom curve (blue) corresponds to the static vertex ($\alpha=0$). Including the frequency structure of the vertex by increasing $\alpha$ changes the dominant superconducting fluctuations from $d$-wave to odd-frequency triplet $s$-wave.
}
\label{Fig_lambda_of_slope}
\end{figure}

\subsubsection{The particle-particle vertex in momentum and real space}

In  Fig.~\ref{Fig_vect_n085} we show the momentum (upper row) and real space $(x,y)$ (lower row) dependence of the $pp$ irreducible vertex  $\Gamma_{pp\uparrow\downarrow}$ for $n=0.85$  obtained on a $6\times 6$ grid and two different values of $U$ in D$\Gamma$A and PA. It is shown for the  bosonic (transfer) momentum of ${\bf q}=(0,0)$ and the lowest Matsubara frequencies, where the contribution is the biggest. The $pp$ irreducible vertex can be interpreted as an effective interaction between electrons that enters the Eliashberg Eq.~\eqref{Eliashberg}. In the momentum space the $pp$ vertex is strongly peaked at  $\mathbf{k}-\mathbf{k'}=(\pi,\pi)$, which is also the maximum 
$\bf{q}$ 
of the magnetic susceptibility. The Fourier transform into real space (lower row of Fig.~\ref{Fig_vect_n085}) corresponds to an attraction between electrons on  neighboring sites, typical for the $d_{x^2-y^2}$-wave symmetry. The full pattern in the real space of the $pp$ vertex is the same in the PA and  D$\Gamma$A  and in agreement with the one obtained in DCA.~\cite{Scalapino_RMP}. There is a strong attractive (negative) interaction between nearest neighbors as well as strong  repulsive (positive) interaction on-site and between second-nearest neighbors  (see Fig.~\ref{Fig_vect_n085}).

For $U=4t$ the momentum and real-space dependence of $\Gamma_{pp\uparrow\downarrow}$ is not only qualitatively but also quantitatively almost the same in  D$\Gamma$A and PA (cf. left and middle column of Fig.~\ref{Fig_vect_n085}). In D$\Gamma$A the magnitude of the nearest-neighbor attraction is slightly lower, which is also reflected in the  $d_{x^2-y^2}$-wave eigenvalue being smaller (cf. Fig.~\ref{Fig_lambda_of_n}). This leads us again to a conclusion that for this value of $U$, there is overall agreement between the PA and D$\Gamma$A, and including the fully irreducible vertex dynamics leads only to a slight suppression of $pp$ fluctuations. 

For  $U=5t$ we present only the PA result (right column of Fig.~\ref{Fig_vect_n085}) since as discussed earlier, no convergent D$\Gamma$A results in this case are available at the moment. Although the temperature is slightly higher than in the $U=4t$ case ($\beta t =15$ vs. $\beta t =20$), the $pp$ vertex is significantly larger and the effective attraction between nearest neighbors almost three times stronger. The overall structure in momentum and real space is however the same as in the case of $U=4t$.

 \begin{figure*}
\centering
\begin{minipage}{\textwidth}
\includegraphics[width=\textwidth]{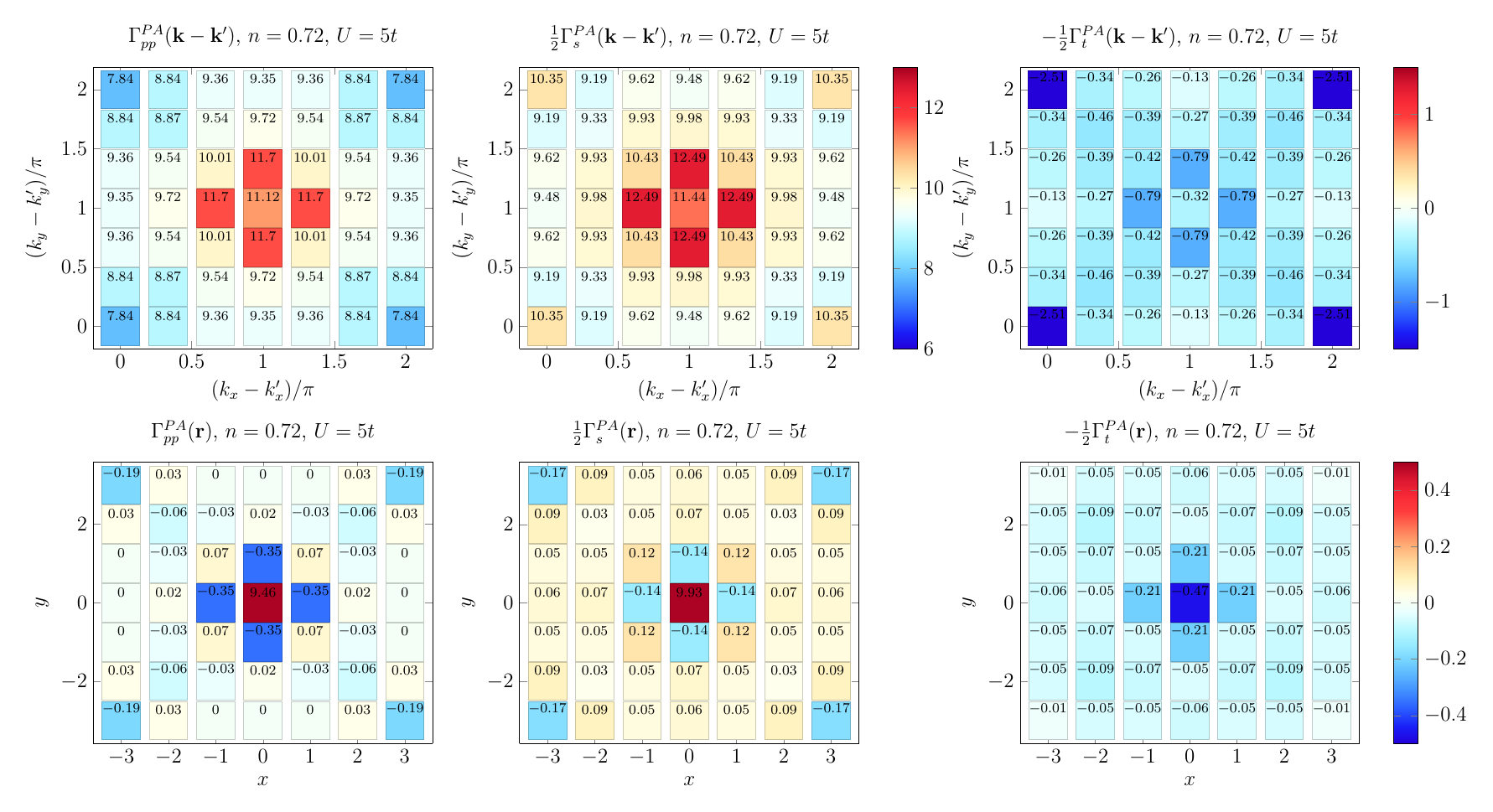}
\caption{Left column: Particle-particle irreducible vertex $\Gamma_{pp\uparrow\downarrow}=\frac{1}{2}(\Gamma_s-\Gamma_t)$ for $U=5t$, $\beta t=15$ and  $n=0.72$  obtained from  PA for a $6\times6$ cluster. The middle and right column show the singlet and triplet contributions, respectively. Upper row: Vertex in momentum space $\Gamma_{pp/s/t}({\bf k}-{\bf k}')$ for $\omega=0$, ${\bf q}=(0,0)$ and $\nu=-\nu'=\pi/\beta$. Lower row:  Fourier transform of $\Gamma_{pp/s/t}({\bf k}-{\bf k}')$ to real space ($x$,$y$). Note: all three plots in the lower row have the same false color scale.}  
\label{Fig_vect_n072_pa}

\end{minipage}

\end{figure*}

 \begin{figure*}
\centering
\begin{minipage}{\textwidth}
\includegraphics[width=\textwidth]{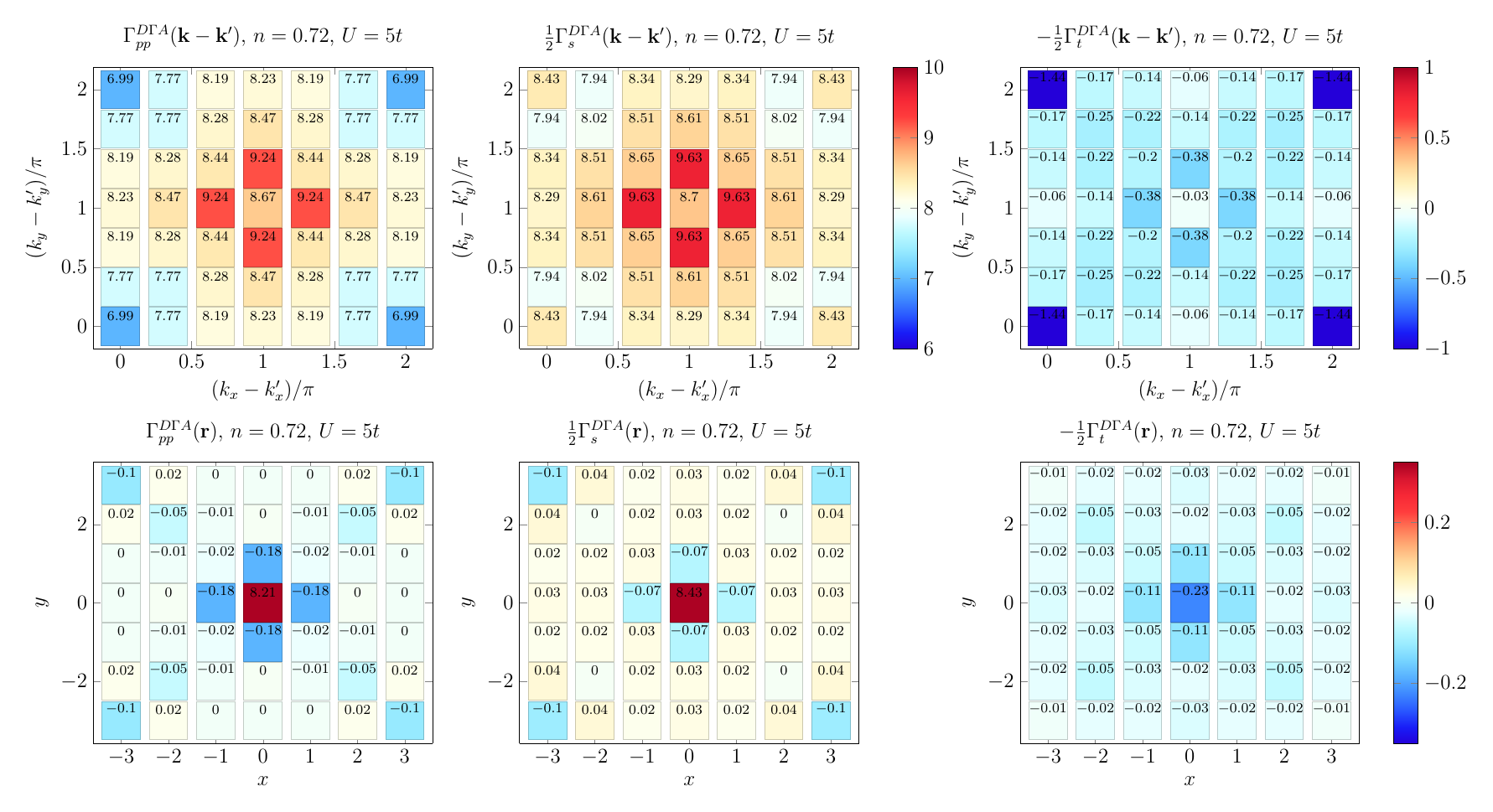}
\caption{Left column: Particle-particle irreducible vertex $\Gamma_{pp\uparrow\downarrow}=\frac{1}{2}(\Gamma_s-\Gamma_t)$ for $U=5t$, $\beta t=15$ and  $n=0.72$  obtained from D$\Gamma$A for a $6\times6$ cluster. The middle and right column show the singlet and triplet contributions, respectively. Upper row: Vertex in momentum space $\Gamma_{pp/s/t}({\bf k}-{\bf k}')$ for $\omega=0$, ${\bf q}=(0,0)$ and $\nu=-\nu'=\pi/\beta$. Lower row:  Fourier transform of $\Gamma_{pp/s/t}({\bf k}-{\bf k}')$ to real space ($x$,$y$). Note: all three plots in the lower row have the same false color scale.} 
\label{Fig_vect_n072_dga}

\end{minipage}

\end{figure*}

\subsection{Eigenvectors: triplet $s$-wave pairing}
\label{Sec:swave}

For higher dopings, the dominant eigenvalue corresponds to a different eigenvector, with a triplet spin symmetry.  As for the case of the $d_{x^2-y^2}$-wave symmetry in Sec.~\ref{dwave}, we present in Fig.~\ref{Fig_susc_n072} the dominant eigenvector $v_s$ in momentum space (left-middle column) together with its projection on the Fermi surface (right-middle column), its Fourier transform into real space (right column) and the respective magnetic susceptibility (left column) from PA (upper row) and D$\Gamma$A (middle row) at $n=0.72$, $\beta t =15$ and $U=5t$ for the $6\times 6$ cluster. For D$\Gamma$A  we also present data for the same parameters but for the $8\times 8$ cluster. At the specific filling of $n=0.72$, in D$\Gamma$A the triplet $s$-wave eigenvector corresponds to the biggest eigenvalue in the $pp$ channel, whereas in the PA this is the second biggest eigenvalue with the $d_{x^2-y^2}$-wave being still the largest one (cf. Fig~\ref{Fig_lambda_of_n}). As we have discussed in Sec.~\ref{dwave}, the ${\bf q}$-dependence of the susceptibility can  help us build an intuitive picture. For both D$\Gamma$A and PA the magnetic susceptibility is strongly peaked at the incommensurate wave vector ${\bf q}=(\pi,\pi-\delta)$ (with $\delta$ dependent on the momentum grid used for the calculation). The value at ${\bf q}=(\pi,\pi)$  is  however still quite large in the case of PA, whereas for D$\Gamma$A it is much smaller (cf. left column of Fig.~\ref{Fig_susc_n072}). The antiferromagnetic fluctuations in PA are thus still strong enough to make  $d_{x^2-y^2}$-wave symmetry in the $pp$ channel the dominant one, whereas in D$\Gamma$A, triplet $s$-wave  {pairing fluctuations} already prevail.

Let us focus now on the triplet $s$-wave eigenvector. Although the projection on the Fermi surface of the eigenvector (right-middle column of Fig.~\ref{Fig_susc_n072}) is very similar in all the three cases shown, the momentum structure and therefore also the real space structure of the eigenvector is slightly different between PA and D$\Gamma$A. Both are $s$-wave (there are no sign changes for the neighbors equally distant from center), but the eigenvector turns negative for the second-nearest neighbors in PA, whereas it stays positive for both cluster sizes in D$\Gamma$A.

\subsubsection{Relation to magnetic susceptibility}

Looking at the peak in magnetic susceptibility $\chi_m(\mathbf{q},\omega=0)$ at ${\bf q}=(\pi,\pi-\delta)$ (shown in the left column of Fig.~\ref{Fig_susc_n072}) and performing the same analysis as was done in Sec.~\ref{dwave} for the possible symmetry of the order parameter,
 we could conclude that the dominant eigenvector should look differently. In particular we could expect sign changes between the points on the Fermi surface connected by the now incommensurate wave vector ${\bf q}=(\pi,\pi-\delta)$ of the (maximal) magnetic  susceptibility. Since with an incommensurate ${\bf q}$ there are 4 maxima in the susceptibility $\chi_m(\mathbf{q},\omega=0)$ in the first Brillouin zone, there would have to be many sign changes of the order parameter on the Fermi surface. Such eigenvector is possible, but we found that the corresponding eigenvalue is significantly smaller than the triplet $s$-wave one. Physically many sign changes would mean many zeroes of the order parameter on the Fermi surface, which is energetically not favorable. The $s$-wave, with no sign changes, seems to be more favorable, although it does not take advantage  {specifically} from the strong peak in the magnetic susceptibility at  ${\bf q}=(\pi,\pi-\delta)$  {but rather from the ${\bf q}$-averaged value. This can be seen from Eliashberg Eq.~\eqref{Eliashberg}: if we approximate the pairing vertex by a product of a frequency dependent part and momentum dependent part, the $s$-wave order parameter without sign changes will pick up the momentum averaged vertex. Similar analysis how the momentum structure and/or frequency structure of the vertex in Eliashberg or eigenvalue equation influences the resulting order parameter can be found in Refs.~\onlinecite{Vojta1999,Tanaka2009, Kitatani2017} }


\subsubsection{Importance of the frequency structure of the $pp$ vertex}
\label{sec_odd_frequency}

In the case where momentum structure cannot be used advantageously enough as explained in the previous paragraph, another factor plays an important role, namely the fermionic frequency dependence of the $pp$ vertex. As can be seen in  Fig.~\ref{Fig_gamma_n072}, (b-c) for the PA and (e-f) for the D$\Gamma$A, the $pp$ vertex is the strongest on the antidiagonal $\nu=-\nu'$ and it increases for larger frequencies.  {As we will see below}, it is this structure that is advantageous for the odd-frequency pairing. With the order parameter  {in the Eliashberg Eq.~\eqref{Eliashberg}} going to zero at $\nu=0$ (necessary in case of $s$-wave triplet pairing) and changing sign,  {these strong contributions at finite frequencies $\nu=-\nu'$ can be picked up.}

Odd-frequency, $s$-wave pairing has, to the best of our knowledge, not been seen before to be the dominant eigenvalue in the 2D Hubbard model without frustration, but it has been reported for more exotic Hamiltonians \cite{Tanaka2008,Tanaka2009,Tanaka2012,Balatsky_review}. Let us emphasize, however, that the methods mainly used in this parameter regime (FLEX, RPA) do not fully consider the fermionic frequency dependence of the $pp$ vertex. 

To check the importance of this frequency dependence, we have used the eigenvalue equation \eqref{EV_eq} to generate eigenvalues and eigenvectors from an interpolative $pp$ vertex defined as
\begin{align}\label{interp_gamma}
&\Gamma_{pp,{\up\dn}}^{interp}(\mathbf{k}, \mathbf{k'},\nu,\nu')=\;\nonumber\\
&\alpha\;[\Gamma_{pp,{\up\dn}}^{D\Gamma A}(\mathbf{k}, \mathbf{k'},\mathbf{0},\nu,\nu',0)
-C_{ {\text{sgn}(\nu),\text{sgn}(\nu')}}(\mathbf{k}, \mathbf{k'})]\nonumber\\
&+C_{ {\text{sgn}(\nu),\text{sgn}(\nu')}}(\mathbf{k}, \mathbf{k'})
\end{align}
where $C_{ {\text{sgn}(\nu),\text{sgn}(\nu')}}(\mathbf{k}, \mathbf{k'})=
\Gamma_{pp,{\up\dn}}^{D\Gamma A}(\mathbf{k}, \mathbf{k'},\mathbf{0},\nu_0,\nu'_0,0)$, is the vertex at the  first Matsubara frequencies 
$\nu_0=\text{sign}(\nu)\frac{\pi}{\beta}$ and $\nu'_0=\text{sgn}(\nu')\frac{\pi}{\beta}$. This way, we interpolate between a frequency independent vertex  {(within the four quadrants given by the sign of $\nu$ and $\nu'$)} which has the same momentum structure ($\alpha=0$) and the full vertex  ($\alpha=1$).
Since we are interested in the effect of the fermionic frequency dependence in the $pp$ vertex, we have considered here only the eigenvalues for the case $\omega=0$ and $\mathbf{q}=0$ and used the D$\Gamma$A Green's functions in Eq.~\eqref{EV_eq}. 

 In Fig.~\ref{Fig_lambda_of_slope} we show the fermionic frequency dependence of  $\Gamma_{pp,{\up\dn}}^{interp}$ for  {the slice} $\nu=-\nu'$ (on the antidiagonal) and for different values of $\alpha$ for one specific momentum point $\mathbf{k}=(\pi,\frac{2}{3}\pi)$, $\mathbf{k'}=(0,0)$. Specifically, we have used the D$\Gamma$A vertex Fig.~\ref{Fig_gamma_n072} (e-f) for which the ratio between the triplet $s$-wave eigenvalue to the $d_{x^2-y^2}$ one is $\lambda_s/\lambda_d=1.63$ and  flattened  the frequency dependence by reducing $\alpha$ in Eq.~\eqref{interp_gamma}, see Fig.~\ref{Fig_lambda_of_slope}.
As we see  from the legend of Fig.~\ref{Fig_lambda_of_slope},
 the $\lambda_s/\lambda_d$ ratio decreases 
when the frequency structure is flattened, down to
 $\lambda_s/\lambda_d=0.81$ for $\alpha=0$. That is, without the frequency dependence of the vertex, $d_{x^2-y^2}$-wave symmetry (singlet, even-frequency) should dominate for this filling, which complies with the RPA results~\cite{Maier_RPA}. 

Note that the values of $\lambda_d$ and $\lambda_s$ obtained from Eq.~\eqref{EV_eq} with the interpolative vertex for different values of $\alpha$ cannot be directly compared since the interpolative vertex is not an actual solution of the problem in any specific approximation. The above analysis  nonetheless indicates that it is the fermionic frequency dependence of the vertex which is important for making odd-frequency $s$-wave pairing the dominant superconducting fluctuation. The odd-frequency $s$-wave competes with $d$-wave even without imposing a strong frustration.


\subsubsection{The particle-particle vertex in momentum and real space}

The particle-particle irreducible vertex $\Gamma_{pp\uparrow\downarrow}$ for $U=5t$, $\beta t=15$ and  $n=0.72$ together with its Fourier transform into real space are shown in Fig.~\ref{Fig_vect_n072_pa} for PA  and in Fig.~\ref{Fig_vect_n072_dga} for D$\Gamma$A for the $6\times6$ cluster. In the three columns of both figures we show the $\Gamma_{pp\uparrow\downarrow}=\frac{1}{2}(\Gamma_s-\Gamma_t)$ vertex (left column) and the singlet $\Gamma_s$ (middle column) and triplet $\Gamma_s$ (right column) contributions. The upper row shows the vertices in momentum space and the lower row their Fourier transform into real space, with the center denoting the on-site (local) contribution. 

Let us first focus on the left column, i.e. $\Gamma_{pp\uparrow\downarrow}$. The momentum  dependence in both PA and D$\Gamma$A  (upper left plot in Figs.~\ref{Fig_vect_n072_pa} and~\ref{Fig_vect_n072_dga})  is peaked at  $(\bf {k-k'})=(\pi,\frac{2}{3}\pi)$  and the three other symmetry related points, which shows again a clear relation between  $\Gamma_{pp\uparrow\downarrow}(\bf{k-k'})$ and the magnetic susceptibility $\chi_m(\mathbf{q},\omega=0)$. In PA the value at  $(\bf {k-k'})=(\pi,\pi)$ is almost as large as at  $(\bf {k-k'})=(\pi,\frac{2}{3}\pi)$ and the structure of the $pp$ vertex in real space (lower left plot in Fig.~\ref{Fig_vect_n072_pa}) resembles the typical one for $d_{x^2-y^2}$-wave pairing symmetry, i.e., nearest-neighbor attraction and second-nearest-neighbor repulsion. Qualitatively the same nearest-neighbor attraction and second-nearest-neighbor repulsion is found for the clear-cut $d$-wave case $n=0.85$ in  Fig.~\ref{Fig_vect_n085} (all lower panels). Quantitatively,
 this  second-nearest-neighbor repulsion is  however very weak. That is, although the dominant eigenvalue for $n=0.72$ in PA corresponds to $d_{x^2-y^2}$-wave symmetry (cf. Fig.~\ref{Fig_lambda_of_n}), the influence of the second largest eigenvalue ($s$-wave) is already visible. 

In  D$\Gamma$A for  the same $n=0.72$, on the other hand,  the $pp$ vertex is not any more large at  $(\bf {k-k'})=(\pi,\pi)$ and the real space structure shown in lower left plot of Fig.~\ref{Fig_vect_n072_dga} differs from the one in PA: The nearest-neighbor attraction is now much weaker and there is an attraction instead of repulsion for the second-nearest neighbors (the vertex is slightly negative also for the third and fourth layer of neighbors). The influence of the $s$-wave contribution is clearly visible. It is also reflected in the $s$-wave eigenvalue being bigger than $d_{x^2-y^2}$-wave (cf. Fig.~\ref{Fig_lambda_of_n}).   

Next, let us look at the singlet vs. triplet contributions shown in middle and right column of Figs.~\ref{Fig_vect_n072_pa} and~\ref{Fig_vect_n072_dga}. For both PA and D$\Gamma$A, the nearest-neighbor attraction is present in the singlet channel, although much weaker in D$\Gamma$A, whereas all further neighbor attraction is entirely in the triplet channel. The negative value of the singlet vertex for the furthest possible neighbors on the diagonal (both in PA and D$\Gamma$A) could either be a higher order $s$-wave contribution, which would be consistent with the incommensurate magnetic ordering vector, or a finite size effect which we cannot eliminate at the moment. This slight furthest neighbor attraction in the singlet channel is also present in the $8\times 8$ cluster calculations for large dopings (not shown here).


\begin{figure}
\centering
\includegraphics[width=0.45\textwidth]{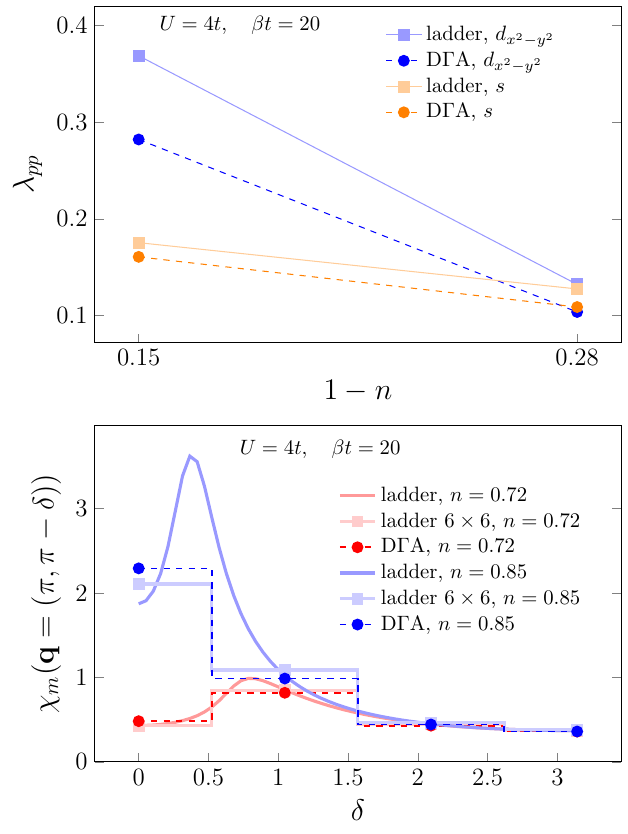}
\caption{Top: comparison of $pp$-eigenvalues for singlet $d_{x^2-y^2}$-wave (blue) and triplet $s$-wave (orange) symmetry between ladder D$\Gamma$A (denoted as 'ladder'; squares) and  parquet D$\Gamma$A on a $6\times 6$ momentum cluster (denoted as 'D$\Gamma$A' as in the previous sections; circles) as a function of doping at $U=4t$ and $\beta t =20$.  Bottom: magnetic susceptibility at $\omega=0$ as a function of $\bf{q}=(\pi,\pi-\delta)$  as obtained in ladder D$\Gamma$A (full lines) and parquet D$\Gamma$A (circles) for two different dopings $n=0.72$ (red) and $n=0.85$ (blue). In addition the results for a ladder D$\Gamma$A calculation on a $6\times 6$ momentum grid are shown (denoted as 'ladder $6\times 6$'; squares).  
}
\label{Fig_comp_ladder_u4}
\end{figure}


\subsection{Comparison with ladder-D$\Gamma$A}
\label{sec:ladder}

In the following we compare the parquet D$\Gamma$A results with the results obtained within the $\lambda$-corrected ladder D$\Gamma$A method, as described in Ref.~\onlinecite{Kitatani2018}. In the ladder method the $pp$ vertices are calculated from $ph$ vertices which in turn are obtained from a Bethe-Salpeter ladder built in the  $ph$  channel (starting from a purely local $\Gamma_m$ and $\Gamma_d$).\cite{footnote4} Additionally, in the calculation of susceptibilities and ladder vertices a Moriasque $\lambda$-correction~\cite{Katanin2009,RMPVertex} is made, which imitates the influence of insertions of $pp$-diagrams that are not present in the ladder. Therefore, the ladder method does not treat the $ph$ and $pp$ channels on equal footing. The advantage of ladder D$\Gamma$A lies in the much larger cluster sizes that it can handle, since the ladder calculations involve only one momentum (contrary to the three needed in the parquet approach).

\begin{figure}
\centering
\includegraphics[width=0.49\textwidth]{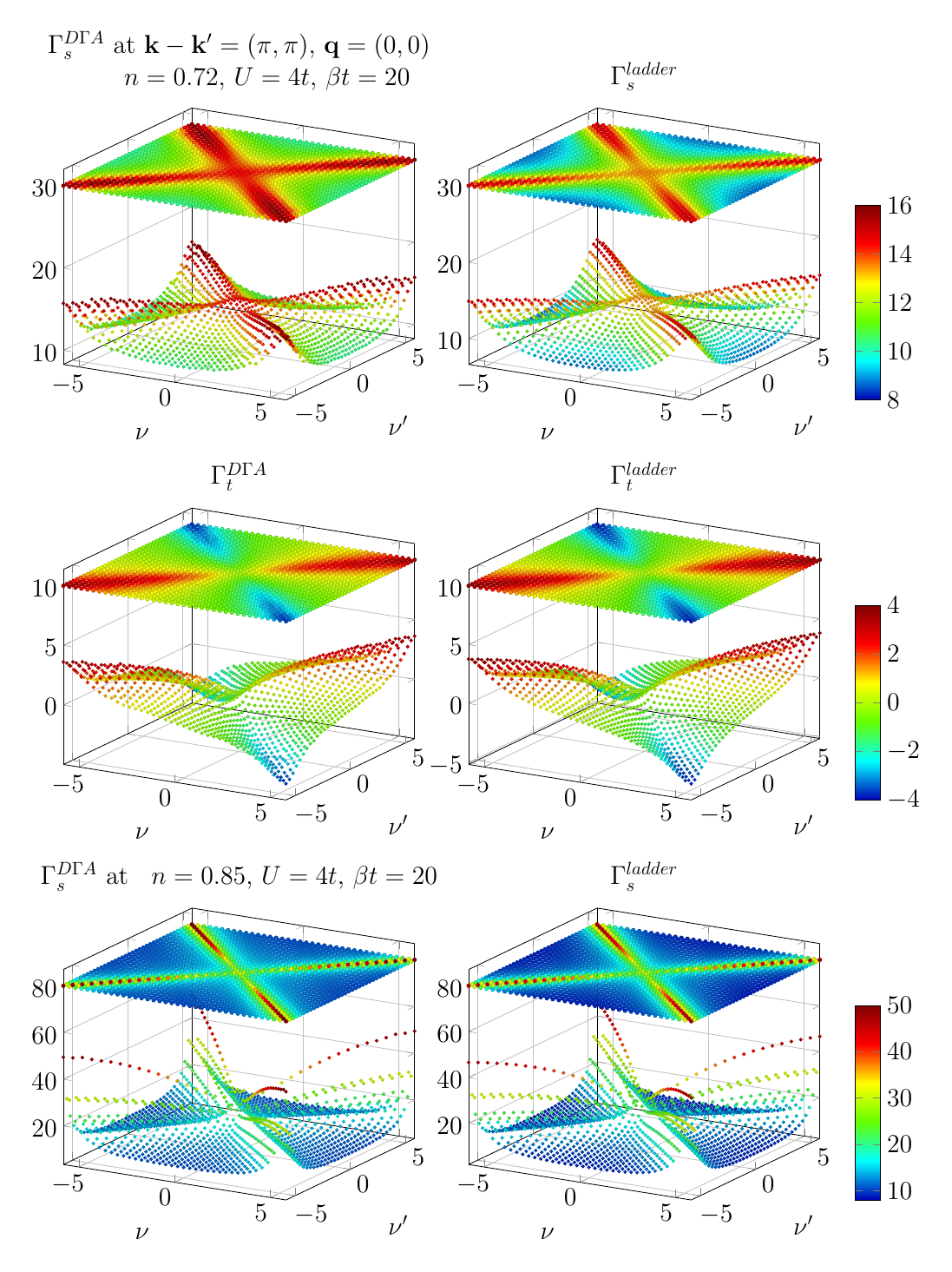}
\caption{ Fermionic frequency dependence of the real part of the irreducible vertex in the $pp$ singlet ($\Gamma^{\nu,\nu',\omega=0}_s$, top) and triplet ($\Gamma^{\nu,\nu',\omega=0}_t$, middle) channels as obtained in the parquet D$\Gamma$A on the $6\times 6$ momentum cluster (denoted as 'D$\Gamma$A', left column) and ladder D$\Gamma$A~\cite{footnote4} (denoted as 'ladder', right column) at $n=0.72$, $U=4t$, $\beta t=20$. Bottom: the filling of $n=0.85$ for the same parameters; shown only for the singlet channel, which is dominating for this filling. The momenta are the same for all plots: ${\bf k}-{\bf k}'=(\pi,\pi)$, ${\bf q}=(0,0)$.}
\label{Fig_comp_gamma_n072_u4}
\end{figure}

In Fig.~\ref{Fig_comp_ladder_u4} the comparison of magnetic and $pp$ eigenvalues (top) and the momentum dependence of static magnetic susceptibilities (bottom) between ladder (denoted as 'ladder') and parquet (denoted as in the rest of the paper 'D$\Gamma$A') versions of D$\Gamma$A is shown. The results are compared for the two different fillings ($n=0.85$ and $n=0.72$), $U=4t$ and $\beta t =20$. For larger doping the difference between the results of the two methods is small and both the eigenvalues and magnetic susceptibility are similar. The eigenvalues are slightly larger in the ladder approximation and also the $pp$ irreducible vertices shown in Fig.~\ref{Fig_comp_gamma_n072_u4} (top two panels) are slightly larger. 

In the case of $n=0.85$ there is a large difference both in $d$-wave eigenvalue and in the susceptibility. There are two reasons for the differences: (i) the insertions from the $pp$ channel are important and are not correctly reproduced by using the  $\lambda$-correction and (ii) finite size effects are large, since the ladder calculation is converged in momentum grid and the parquet one used here is on the $6\times 6$ cluster.

In order to get some insight which of the effects dominates we have calculated the ladder D$\Gamma$A magnetic susceptibility on a $6\times6$ cluster (shown in Fig.~\ref{Fig_comp_ladder_u4} as the light blue for $n=0.85$ and pink for $n=0.72$). For $n=0.85$ the difference is quite dramatic, since on the $6\times 6$ cluster the incommensurate peak in the susceptibility is missing and its weight is distributed  between the $(\pi,\pi)$ and $(\pi,\frac{2}{3}\pi)$ points. That could explain the large difference of the $d$-wave eigenvalues in the two methods and supports the (ii) explanation for the difference in the two methods. At higher temperatures, when the momentum resolution is less important, the eigenvalues lie closer to each other as can be seen in the upper panel of Fig.~\ref{Fig_lambda_T}.

The $pp$ vertices obtained in the two methods for the same parameters as above, shown in   Fig.~\ref{Fig_comp_gamma_n072_u4} also look similar, even for the filling of $n=0.85$, which lets us conclude that the $\lambda$-correction in the ladder method fairly well reproduces the influence of diagrams in the $pp$ channel for these parameters.   Instead it is the finite grid which, for particular dopings, might miss the proper incommensurate wave vector.

\begin{figure}
\centering
\includegraphics[width=0.49\textwidth]{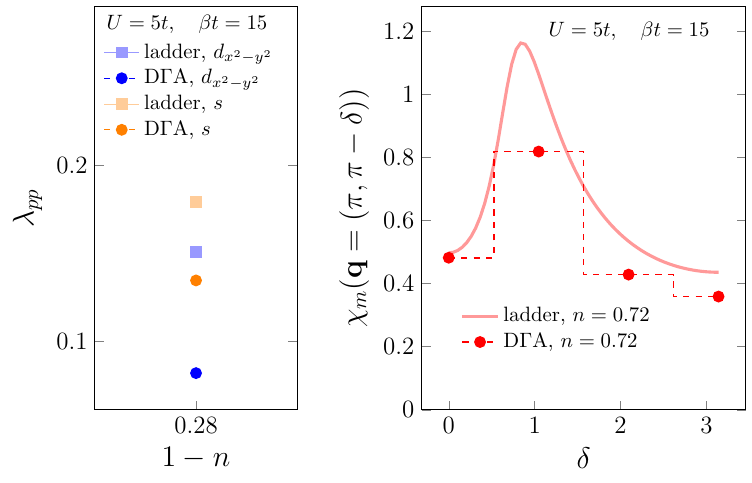}
\caption{The same as Fig.~\ref{Fig_comp_ladder_u4} for doping $n=0.72$ and $U=5t$, $\beta t=15$ }
\label{Fig_comp_ladder_u5}
\end{figure}

\begin{figure}
\centering
\includegraphics[width=0.49\textwidth]{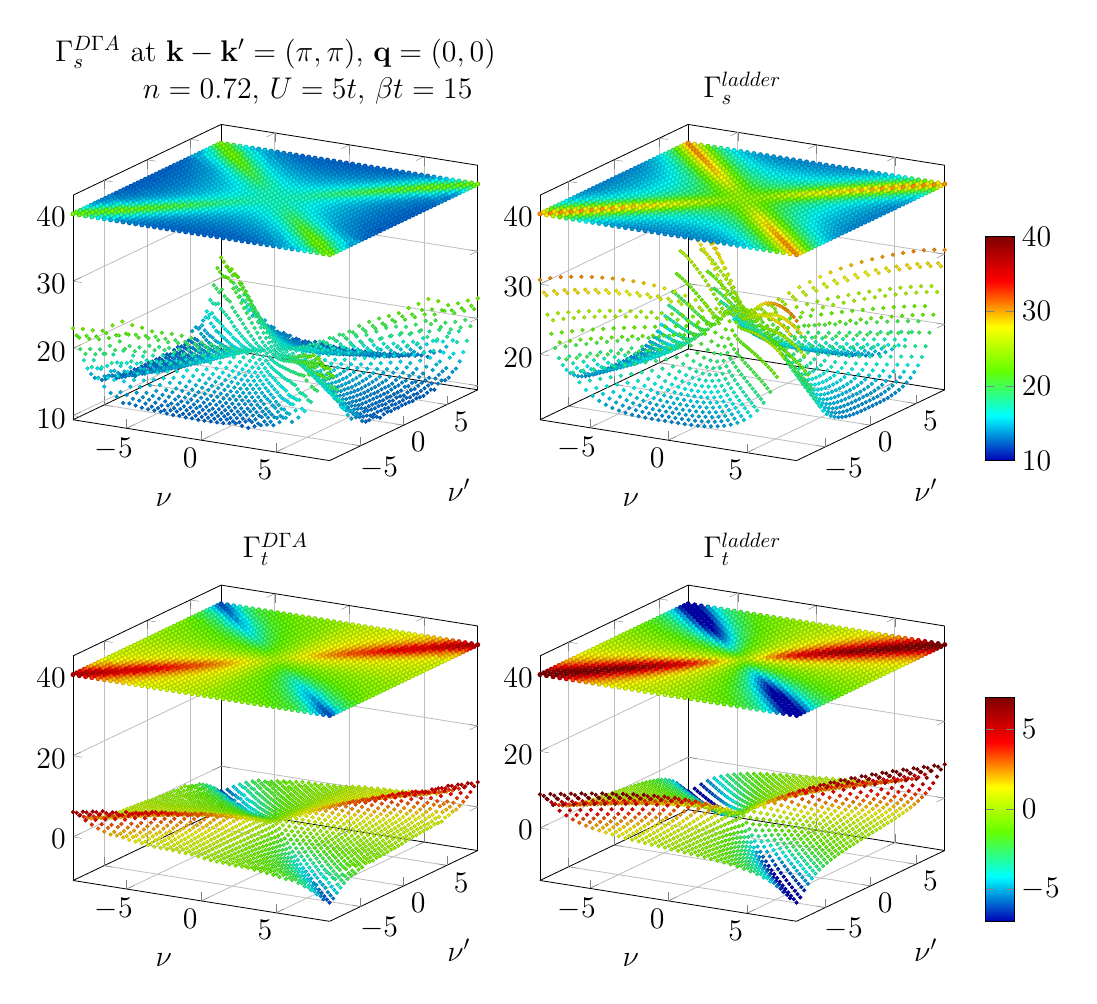}
\caption{Fermionic frequency dependence of the real part of the irreducible vertex in the $pp$ singlet ($\Gamma^{\nu,\nu',\omega=0}_s$, top) and triplet ($\Gamma^{\nu,\nu',\omega=0}_t$, bottom) channels as obtained in the parquet D$\Gamma$A on the $6\times 6$ momentum cluster (denoted as 'D$\Gamma$A', left column) and ladder D$\Gamma$A~\cite{footnote4} (denoted as 'ladder', right column) at $n=0.72$, $U=5t$, $\beta t=15$. The momenta are the same for all plots: ${\bf k}-{\bf k}'=(\pi,\pi)$, ${\bf q}=(0,0)$.}
\label{Fig_comp_gamma_n072_u5}
\end{figure}

In Fig.~\ref{Fig_comp_ladder_u5} the results analogous to those in Fig.~\ref{Fig_comp_ladder_u4} are shown for $U=5t$, $\beta t=15$ and only one filling of $n=0.72$. The ladder eigenvalues also show the triplet $s$-wave as the dominant symmetry in the $pp$ channel and they are both bigger than the parquet eigenvalues. Here the reasoning that eigenvalues could be smaller in the parquet D$\Gamma$A due to missing of a strong peak in the magnetic channel is not valid, mainly because the peak is not so pronounced and also for odd frequency $s$-wave pairing the momentum dependence is not that important as the frequency dependence as explained in Sec.~\ref{sec_odd_frequency}. Looking at the frequency dependence of the $pp$ vertices in both methods shown in  Fig.~\ref{Fig_comp_gamma_n072_u5} we see that the  $pp$ vertices in the ladder method  are much stronger than in parquet D$\Gamma$A, which leads to larger eigenvalues. Also the frequency dependence of the ladder vertices is steeper which leads to a larger ratio  ${\lambda_s}/{\lambda_d}$ of the eigenvalues, compliant with the discussion in Sec.~\ref{sec_odd_frequency}.

\section{Conclusion}
\label{Sec:Summary}
 {We have performed a parquet equations study of the doped two dimensional 
Hubbard model using two different approximations for the fully irreducible vertex: the PA and parquet D$\Gamma$A}. In contrast to previous diagrammatic extensions of DMFT,\cite{Otsuki2014,Kitatani2018,Vucicevic2017,RMPVertex} we employ the parquet equations instead of using ladder diagrams  for generating non-local correlations. Hence our approach is not biased against a certain  {scattering} channel. All relevant fluctuations for unconventional superconductors, i.e. magnetic, superconducting, charge density wave fluctuations, are treated on an equal footing. This is particularly important since several instabilities are at close quarters  {\em and}  mutually influence each other:
commensurate and incommensurate antiferromagnetism, singlet $d$-wave and triplet $s$-wave pairing.

Close to half-filling, we find predominant commensurate antiferromagnetic fluctuations at wave vector ${\mathbf q}=(\pi,\pi)$, and the leading  {fluctuation in the pairing channel} are $d$-wave.  At finite but small dopings, e.g., at 15\% doping or $n=0.85$ electrons per site, the $d$-wave  eigenvalue increases much faster 
 upon lowering temperature than the antiferromagnetic one which even decreases at the lowest temperature accessible [Fig.~\ref{Fig_lambda_T} (top)].  {Although we cannot exclude that another instability will start dominating at even lower temperatures, it is likely} that for this doping $d$-wave superconductivity eventually prevails,  {however with a smaller $T_c$ than many other methods predict}.~\cite{Kyung2003,Hafermann2008,Gull2013, Otsuki2014,Kitatani2015} 
From the structure of the eigenvalue (see below) and also from the fact that
 commensurate antiferromagnetic and $d$-wave superconducting fluctuations
show the same trends, e.g., they fall off concomitantly if we move toward higher doping [Fig.~\ref{Fig_lambda_of_n}], we conclude that $d$-wave  {pairing fluctuations} are induced by  {(nearly)} commensurate antiferromagnetic spin fluctuations. Let us also emphasize in this context that in the temperature and doping range studied we do not find charge density wave fluctuations with magnitude comparable to the magnetic ones. 

At larger doping, we change from commensurate to  {highly} incommensurate antiferromagnetic spin fluctuations  with e.g. ${\mathbf q}=(\pi,2\pi/3)$, and from singlet $d$-wave to  triplet $s$-wave  {fluctuations in the pairing channel}. Both eigenvalues,  incommensurate antiferromagnetic and $s$-wave  {pairing}, show a similar doping dependence [Fig.~\ref{Fig_lambda_of_n}].

 If we connect points of the Fermi surface by the commensurate antiferromagnetic
wave vector ${\mathbf q}=(\pi,\pi)$, it becomes clear that commensurate antiferromagnetic and $d$-wave fluctuations match perfectly. 
The latter has a sign change for those points of the Fermi surface connected by  ${\mathbf q}=(\pi,\pi)$, see Fig.~\ref{Fig_susc_n085}.
For  incommensurate antiferromagnetism,  there are now however four equivalent  incommensurate wave vectors within the Brillouin zone [${\mathbf q}=(\pi,2\pi/3)$, ${\mathbf q}=(2\pi/3,\pi)$, ${\mathbf q}=(\pi,-2\pi/3)$, ${\mathbf q}=(-2\pi/3,\pi)$, see Fig.~\ref{Fig_susc_n072}]. A similar line of reasoning as for the commensurate wave vector would hence require many changes of sign on the Fermi surface. Such a complicated superconducting order parameter is conceivable, and a possible solution in our PA or parquet D$\Gamma$A calculation. Indeed we  find it to be  the second largest pairing eigenvalue at large doping; the largest one is however a triplet $s$-wave eigenvalue [shown in  Fig.~\ref{Fig_susc_n072}].  {In the case of triplet odd-frequency $s$-wave pairing, the order parameter does not specificaly profit from a pronounced peak at incommensurate ${\mathbf q}$, but rather from a ${\mathbf q}$-averaged value. Another factor which makes} the  odd-frequency  $s$-wave pairing fluctuations dominate  {over even in frequency $d$-wave} is the strong frequency structure of the vertex $\Gamma_{pp}$ which is suppressed at small fermionic frequencies and away from the line with fermionic frequencies $\nu'=-\nu$.

 Our detailed analysis in real and $\mathbf k$-space does not only help us to unambiguously  identify the eigenvalues  {in the pairing channel}, but it  also reveals the evolution  from $d$- to $s$-wave. In real space, the superconducting pairing glue $\Gamma_{pp}$ always has a strong local repulsion, reflecting the repulsive $U$ of the Hubbard model. For small dopings the antiferromagnetic spin-fluctuations provide however a substantial nearest-neighbor attraction along with second-nearest-neighbor repulsion [Fig.~\ref{Fig_vect_n085} (left)]. This eventually leads to a strong  nearest-neighbor component of the $d$-wave eigenvalue with the typical  sign change  in real space [Fig.~\ref{Fig_susc_n085} (right)]. For larger dopings this nearest-neighbor attraction gets weaker and the second-nearest-neighbor repulsion eventually becomes attractive [Fig.~\ref{Fig_vect_n072_dga} (bottom left)]. Here, the $s$-wave eigenvalue becomes favorable in D$\Gamma$A, which has again a strong  nearest-neighbor component but now without sign changes [Fig.~\ref{Fig_susc_n072} (lower and middle right)].

 {To sum up, our new development of the parquet equations solver enables us to analyze low-temperature physics of the doped Hubbard model and visualize all possible fluctuations in an unsupervised manner. We obtained that for smaller doping levels commensurate antiferromagnetic fluctuations lead to strong $d$-wave pairing fluctuations, which become weaker as the doping is increased. For larger doping levels, 
 we also observed that the triplet $s$-wave odd-frequency pairing fluctuations become comparable (or even dominate) and we raised the frequency structure of the vertex as a new (but general) factor for supporting this observation. Since the temperature is still not low enough for discussing phase transitions, our unbiased study will provide useful insight for future studies on magnetism and superconductivity.}

\subsection*{Acknowledgements}
We thank  J. Kune\v{s},  P. Pudleiner, and T. Ribic for very helpful and inspiring discussions and C. Watzenb\"ock for obtaining some of the D$\Gamma$A input. The help of the Vienna Scientific Cluster (VSC) Team, in particular of C. Blaas-Schenner, in optimizing the code is also greatly acknowledged.
The present work was supported financially by 
European Research Council under the 
European Union's Seventh Framework Program
(FP/2007-2013) through ERC Grant No. 306447
and the Austrian Science Fund (FWF) through SFB ViCoM F41 and project P 30997. G.L. acknowledges the starting grant of ShanghaiTech University, the Program for Professor of Special Appointment (Shanghai Eastern Scholar) and the support from the National Natural Science Foundation of China (Grant No. 11874263).
Calculations have been done
on the VSC.

\appendix
\section{Convergence of the eigenvalues with respect to the number of frequencies}
\label{Apppendix:convergence}

   \begin{figure}
\centering
\includegraphics[width=0.45\textwidth]{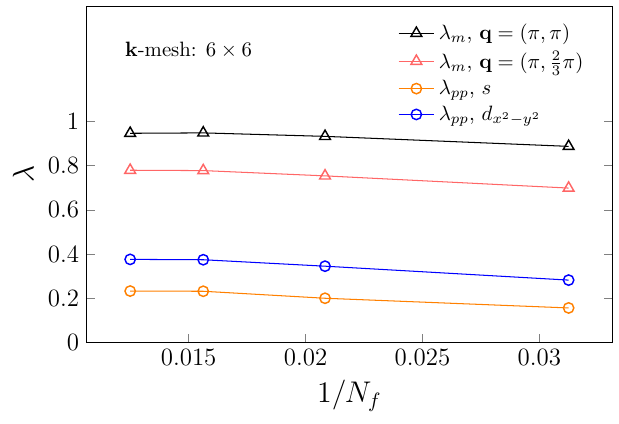}
\caption{The dominant eigenvalues in the magnetic ($\lambda_m$, triangles) and particle-particle ($\lambda_{pp}$, circles) channels as a function of number of frequencies $N_f$ used for the calculation. The momentum cluster size is $6\times 6$ and the parameters are $U=5t$, $n=0.85$, $\beta t=15$.}
\label{Fig_lambda_of_size}
\end{figure}

The parquet equations solver scales with the power of 3.4  in the number of frequencies and momenta used (stemming from one prefactor [power 1] in the number of frequencies/momenta and from a matrix-matrix multiplication [power 2.4]; using the point group symmetry reduces only the prefactor of this scaling). The {\it victory} code actual efficiency is however memory bound (see Ref.~\onlinecite{victory}). For the $8 \times 8$ momentum cluster most of the calculations were done for 64 frequencies (32 positive and 32 negative) and it was not possible to use more frequencies. At the lowest temperatures presented  for the $8 \times 8$ cluster ($\beta t =10$ for $U=5t$ shown in Fig.~\ref{Fig_lambda_T}, as well as $\beta t =15$ for $U=5t$ and  $\beta t =20$ for $U=4t$ in Figs.~\ref{Fig_susc_n072} and~\ref{Fig_susc_n085}) we were not able to obtain convergent results for a smaller number of frequencies. Therefore, $8 \times 8$ cluster results at these temperatures are not converged with respect to the number of frequencies. Convergence analysis was done for the $6\times 6$ case, however. The frequency-box-size dependence of the eiganvalues for $\beta t=15$, $U=5t$ is shown in Fig~\ref{Fig_lambda_of_size}. Since the values do not change significantly between the box of 128 frequencies and the largest box taken (160 frequencies), most of the results presented in this paper were obtained using 128 frequencies. For higher temperatures, the results do not change already for 64 frequencies. Based on the results for the $6\times 6$ cluster, we can expect some quantitative differences in the eigenvalues for the $8\times 8$ cluster if we take a larger frequency box. However, we do not expect a qualitative change of the overall behavior.


\begin{thebibliography}{94}
\bibitem [{foo({\natexlab{a}})}]{footnote1}%
  \BibitemOpen
  \href@noop {} {} ({\natexlab{a}}),\ \bibinfo {note} {that is, cutting a
  Green's function line does not separate the Feynman diagram into two
  pieces.}\BibitemShut {Stop}%
\bibitem [{foo({\natexlab{b}})}]{footnote3}%
  \BibitemOpen
  \href@noop {} {} ({\natexlab{b}}),\ \bibinfo {note} {for both $U=4t$ and
  $U=5t$ the temperature is the lowest for which we could converge the PA
  calculations at all fillings.}\BibitemShut {Stop}%
\end{thebibliography}
%

\end{document}